\documentclass[reprint,twocolumn, amsmath,amssymb, aps]{revtex4}

\usepackage{graphicx}
\usepackage{dcolumn}
\usepackage{bm}
\usepackage{ulem}
\usepackage{mathtools}
\usepackage{natbib}
\usepackage{xcolor}
\newcommand{\NBI}{\affiliation{Niels Bohr Institute, University of Copenhagen, Blegdamsvej 17, 2100 Copenhagen, Denmark}}

\newcommand{\HyQ}{\affiliation{Center for Hybrid Quantum Networks (Hy-Q), Niels Bohr Institute, University of Copenhagen, Copenhagen, Denmark}}

\renewcommand{\deg}{\ensuremath{{}^{\circ}}}
\DeclarePairedDelimiter\set\{\}

\begin{document}

\preprint{APS/123-QED}

\title{Soft-clamped phononic dimers for mechanical sensing and transduction}
\author{Letizia Catalini}
\thanks{These authors contributed equally to this work.}

\author{Yeghishe Tsaturyan }
\thanks{These authors contributed equally to this work.}
\altaffiliation{Present address: Pritzker School of Molecular Engineering, University of Chicago, Chicago, IL 60637, USA.}
\email{ytsaturyan@uchicago.edu}

\author{Albert Schliesser}
\email{albert.schliesser@nbi.ku.dk}

\NBI
\HyQ

\begin{abstract}
Coupled micro- and nanomechanical resonators are of significant interest within a number of areas of research, ranging from synchronization, nonlinear dynamics and chaos, to quantum sensing and transduction.
Building upon our work on soft-clamped membrane resonators, here we present a study on \textit{phononic dimers}, consisting of two defects embedded in a phononic crystal membrane.
These devices exhibit widely tunable (2-100~kHz) inter-defect coupling strengths, leading to delocalized hybrid modes with mechanical $Qf$-products $>10^{14}~\mathrm{Hz}$ at room temperature, ensuring low thermomechanical force noise.
The mode splitting exhibits a strong dependence on the dimer orientation within the crystal lattice, as well as the spatial separation between the two defects.
Given the importance of dynamic range for sensing applications, we characterize the relevant mechanical nonlinearities, specifically the self- and cross-Duffing parameters, as well as self- and cross-nonlinear dampings.
This work establishes soft-clamped resonators with engineered spatial and spectral multi-mode structure as a versatile mechanical platform both in the classical and quantum regimes. 
Applications in microwave-to-optical transduction and magnetic resonance force microscopy are particularly attractive prospects.
\end{abstract}

\maketitle

\section{Introduction}
Mechanical resonators are widely used to translate force signals into a displacement, which is subsequently detected using optical or electronic techniques.
A well-known example is the atomic force microscope.
The mechanical system's thermal fluctuations set a lower limit for the amount of noise added in such a cascade.
These fluctuations can be quantified as an equivalent force noise of (double-sided) power spectral density ${S_\mathrm{FF}=2 m \Omega_\mathrm{m} k_{\mathrm{B}} T/Q}$, where $m$ is the resonator mass, $\Omega_\mathrm{m}$ its resonance frequency, $k_\mathrm{B}$ the Boltzmann constant, $T$ the device temperature, and $Q$ the mechanical quality factor.
To achieve high sensitivity, it is therefore imperative to use mechanical devices with low mass  and high quality factor $Q$.

We have recently introduced a novel approach that allows boosting the quality factors available in low-mass nano- and micromechanical resonators by several orders of magnitude, reaching into the regime of $Q\sim10^9$ for MHz-resonance frequencies.
This ``soft-clamping'' approach  \cite{tsaturyan_ultracoherent_2017} is based on phononic engineering, and harnesses periodic structures with an acoustic bandgap to confine the mechanical vibrations.
Importantly, the localization of mechanical vibrations minimizes the bending of the mechanical device, which is known to dominate dissipation in tensioned string and membrane resonators \cite{Gonzalez1994, Unterreithmeier2010, Villanueva2014}.

Due to their extraordinarily low noise and high coherence, soft-clamped mechanical resonators are currently explored in a number of settings \cite{tsaturyan_ultracoherent_2017, Ghadimi_2018, Reetz_2019, Fedoseev2019}.
Applications in which tensioned micro-membrane resonators have previously shown to be promising, such as magnetic force resonance microscopy (MRFM) \cite{Scozzaro2016, Reetz_2019, Fischer2018, Kosata_2020}, and electro-opto-mechanical quantum transducers \cite{Bagci2014, Andrews2014, Higginbotham2018}, can benefit from the high $Q$-factors achieved with soft clamping.

In such applications, it can be important to spatially separate the region in which the force of interest acts on the mechanical mode, and the region in which the induced motion is read out optically.
This can help avoid damage to the structure that exerts the force (a biological specimen, or a superconducting device, for example), by the light used for displacement detection.

Phononic engineering again offers great flexibility in meeting all these demands simultaneously. 
Here, we concretely explore design based on phononic dimers, i.~e.~containing a pair of identical defects to which mechanical modes are localized. 
The defect modes hybridize into a pair of modes, each delocalized over both defects. The principle of dimerization in 2D phononic crystal structures has previously been studied in other devices \cite{Li_2005, Miyashita_2008, Khelif_2003, Lanzillotti_Kimura_2010}.
In the following, we show that significant hybridization --- in terms of coupling strength and therefore frequency splitting --- can be achieved while realizing the desired spatial separation of interaction regions.
We also investigate the quality factors and masses realized in such devices, as well as the nonlinear properties of the modes under large-amplitude drive. Our findings feed into the analysis of parametric spin-sensing protocols \cite{Kosata_2020} based on the phononic force sensors presented here.
%

\section{Device principle and dimerization}

\begin{figure}[htb]
\center
\includegraphics[scale=1]{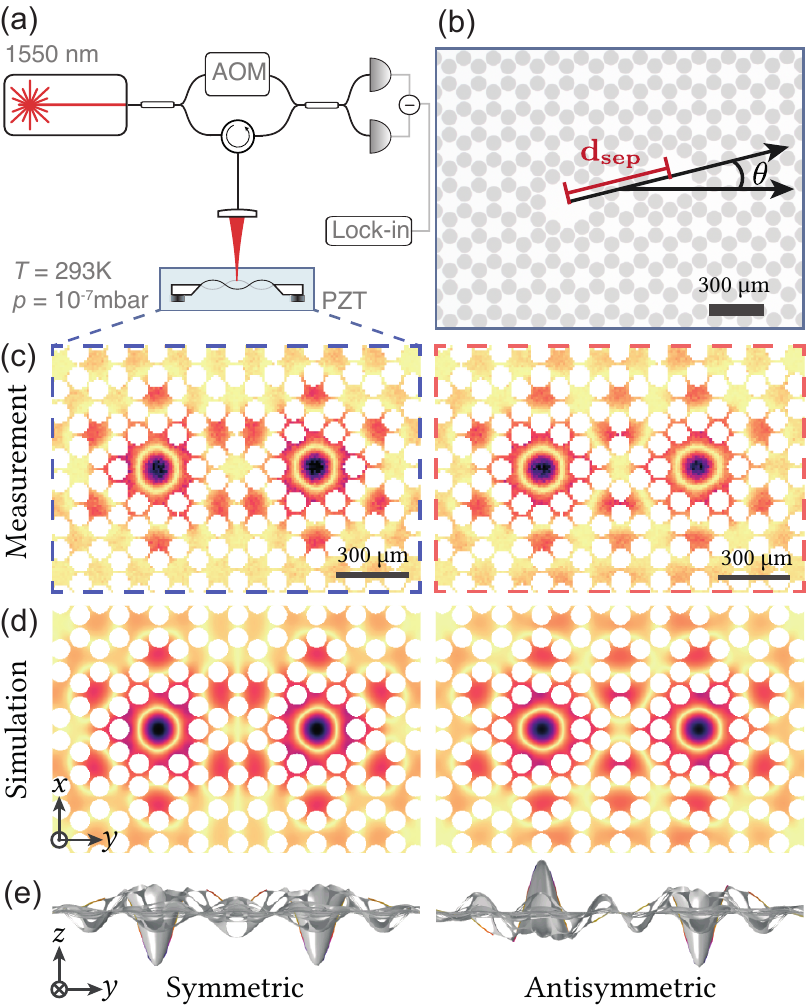}
\caption{
(a) Sketch of the experimental setup. Laser light is reflected from the membrane under test, and its heterodyne beat note with a local oscillator shifted by an acousto-optic modulator (AOM) recorded in a balanced photodetector. The photodetector signal is recorded with a lock-in amplifier. Membrane motion can be excited by means of a piezoelectric actuator (PZT). 
(b) Micrograph of a device, with dimer separation and orientation indicated.
(c) Measured and (d) simulated out-of-plane displacement profiles of the symmetric and antisymmetric modes of a phononic dimer structure. 
(e) Cross-sectional view of the simulated displacement profiles emphasizes the even and odd parity of the modes.}
\label{f:fig1}
\end{figure}

The devices are based on highly stressed, suspended silicon nitride membranes (thickness $h\approx 14~\mathrm{nm}$), which have been patterned with a honeycomb phononic crystal structure (Fig.\ \ref{f:fig1}).
With a lattice constant of 
$a\approx 160~\mathrm{\mu m}$ and hole radii of $r=0.26a$, the crystal structure exhibits a phononic stop-band around $1.4~\mathrm{MHz}$. 
Upon introducing a geometric defect of suitable dimensions within the honeycomb lattice, vibrational modes can be spatially localized within said defect \cite{tsaturyan_ultracoherent_2017}. 
Importantly, and in contrast to the modes of the entire membrane structure, the localized modes are subject to a qualitatively different clamping condition. The evanescent decay into the perforated membrane structure ensures negligible mode amplitude at the points of contact between the membrane and the underlying (silicon) substrate. This results in a significant reduction of the total mode curvature and hence elastic energy dissipation \cite{Schmid2011, Unterreithmeier2010, Yu2012}. The localized modes are soft clamped \cite{tsaturyan_ultracoherent_2017} and devices based on this principle have recently shown quality factors in excess of $10^9$ at moderate cryogenic temperatures \cite{Rossi2018}.

As we introduce a second defect within the crystal lattice, new eigenmodes are formed, constituting the solutions to the eigenvalue problem 
\begin{equation}
    K u = \Omega^2 M u,
\end{equation}
where  $\Omega$ is the eigenfrequency, $u$ is a vector of mode amplitudes $u_n$, {and $M$ ($K$) is the mass (spring) matrix.}
Each amplitude $u_n$ describes the maximum out-of-plane displacement of a membrane mode with a normalized mode shape $\phi_n(x,y)$, so that the total displacement of of the membrane resonator is ${w(x,y)=\sum_n u_n \phi_n(x,y)}$.
This approach is similar in spirit to the linear combination of atomic orbitals in a dimer molecule.
In a membrane {structure} under high in-plane tensile stress $\bar \sigma$, the mass and spring matrices are approximately given by $M_{nm}=\rho \langle \phi_n | \phi_m \rangle $ and $K_{nm}=\bar{\sigma}\langle \nabla\phi_n | \nabla\phi_m \rangle $, respectively, where $\langle\cdot\rangle$ denotes integration over the volume of the resonator (see Appendix \ref{app:lumpedElement} and ref.~\cite{Midtvedt_2014}).

For two sufficiently distant, identical defects within the crystal lattice, the system is well approximated by the basis functions $\phi_n$ of the two ``local'' modes, that is, the radial modes of the first ($\phi_1$), and second ($\phi_2$) defect in absence of the other defect.
As a result we obtain new eigenmodes 
\begin{align}
  \phi_\mathrm{S}(x,y) &\approx \phi_1(x,y)+\phi_2(x,y)
  \label{eq:symmModeSpatial}\\
  \phi_\mathrm{A}(x,y) &\approx \phi_1(x,y)-\phi_2(x,y),
  \label{eq:asymmModeSpatial}
\end{align}
delocalized over both defects.
We refer to them as the symmetric (S) and antisymmetric (A) hybrid modes, with even and odd parity (see Fig. \ref{f:fig1}), respectively. 
To lowest order in the off-diagonal elements, the eigenfrequencies of these modes are split by an amount (see Appendix \ref{app:lumpedElement})
\begin{equation}
    \label{eq:splitting}
      \frac{\Omega_\mathrm{A}-\Omega_\mathrm{S} }{\Omega_0}\approx
      \left(
        \frac{M_{12}}{M_{11}}-\frac{K_{12}}{K_{11}}
      \right),
\end{equation}
with a mean frequency close to the eigenfrequency ${\Omega_0\equiv\sqrt{K_{11}/M_{11}}}$ of the individual defects.

\section{Frequency splittings, damping and modal masses}
The splitting between the antisymmetric and symmetric modes, $|\Omega_\mathrm{A}-\Omega_\mathrm{S}|/2\pi$, is determined by the coupling strength of the modes localized within the two defects.
We can tune the coupling strength by changing the distance between the defects, $d_\mathrm{sep}$, as well as the orientation $\theta$ of the vector that connects the two defects, with respect to the phononic lattice (see Fig. \ref{f:fig1}-b for an illustration).
We have implemented and characterized devices with $\theta=\{0\deg,\,13.9\deg,\,30\deg,\,46.1\deg\}$ and varying distance between the defects.
The distances chosen roughly coincide with 0, 1, 2 and 3 unit cells between the defects.

The membrane resonators are fabricated in a $\sim 14~\mathrm{nm}$ thin silicon nitride film, with an in-plane tensile stress of $\bar{\sigma}\approx 1.27 ~\mathrm{GPa}$ (cf. Appendix \ref{app:fabrication} for further details).
The samples are subsequently characterized using a home-built optical interferometer based on a 1550~nm fiber laser and a heterodyne detection scheme (see Fig.~\ref{f:fig1}-a and Appendix \ref{app:setup} for details).
The experiments are carried out at room temperature, and pressures below $10^{-7}$~mbar.
From the interferometer signal we can extract displacement spectra, quality factors and effective masses.
Furthermore, on a similar setup (see \cite{Barg_2016} for details) we perform spatial raster scans of the membrane devices, acquiring $(x, y)$-maps of the out-of-plane displacement of a particular mode, as shown in Fig.~\ref{f:fig1}-c.

\begin{figure*}[htb]
\center
\includegraphics[width=17.2cm]{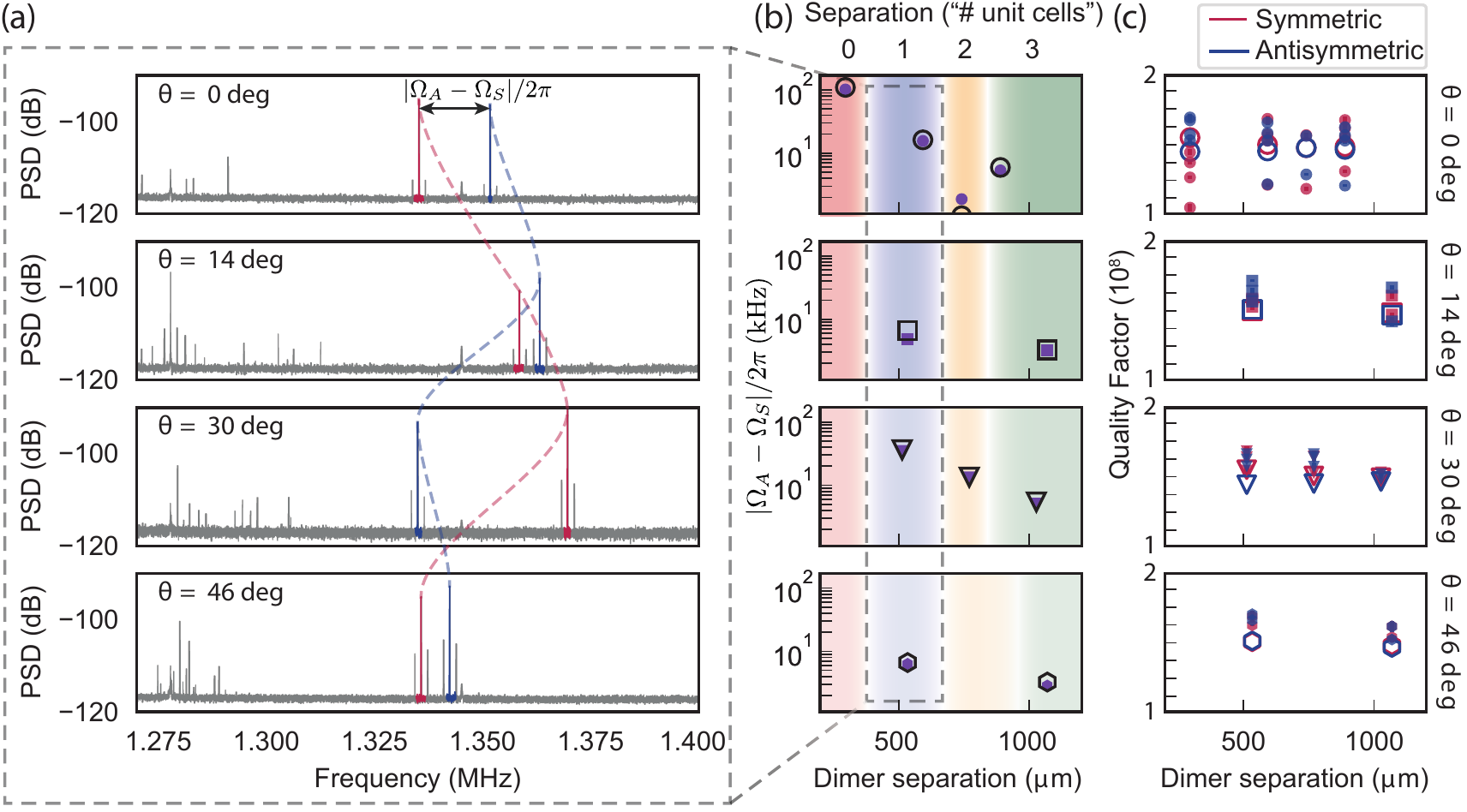}
\caption{
(a) Thermomechanical noise spectra for four different orientations  $\theta$ with $d_\mathrm{sep}=1\mathrm{\,unit\,cell}$, with peaks corresponding to the symmetric (red), and antisymmetric (blue) mode highlighted.
Notice that the order between the two changes for $\theta=30\deg$ (independent of $d_\mathrm{sep}$, and also for $\theta=0\deg$, $d_\mathrm{sep}=3$ unit cell). The dashed lines are a guide to the eye, drawn to highlight the change in modal order as a function of dimer orientation.
(b) Measured frequency splitting for all orientations $\theta$ as a function of dimer separation $d_{\mathrm{sep}}$ (filled symbols). 
Empty symbols indicate simulated frequency splitting.
(c) Measured quality factor for all orientations as a function of dimer separation, for the symmetric (red) and antisymmetric (blue) mode (filled symbols).
Empty symbols indicate simulated $Q$-factor.
}
\label{f:fig2}
\end{figure*}

Upon varying the sample geometry, the frequency splitting between the symmetric and antisymmetric modes changes from $\sim 2\,\mathrm{kHz}$ to $\sim 100\,\mathrm{kHz}$, as shown in Fig \ref{f:fig2}-a.
Importantly, for comparable dimer separations (i.e., $d_\mathrm{sep}\sim 500~\mathrm{\mu m}$) the frequency splitting can vary by almost an order of magnitude, depending on the orientation of the dimer within the honeycomb lattice.
We can compare the obtained frequencies with finite element simulations (see Appendix \ref{app:simulation} for details), and find excellent agreement.
Interestingly, the frequency ordering of symmetric and antisymmetric modes can change with the geometry (see Fig. \ref{f:fig2}c).

Next, we extract the quality factor for the hybrid modes, by exciting them one at a time with a piezoelectric actuator.
After the drive is turned off, we observe the ringdown of the mechanical mode as a function of time using the optical interferometer.
From the characteristic amplitude ringdown time, $2/\Gamma_i$, we calculate the quality factor $Q_i=\Omega_i/\Gamma_i$, which is shown for all measured devices and modes in Fig.~\ref{f:fig2}-b.
For all measured membranes, the quality factors are larger than 100 million.
This is consistent with finite element modeling of dissipation in these soft-clamped devices as shown in Fig.~\ref{f:fig2}-b (see Appendix \ref{app:simulation} for details). 

Finally, we evaluate the effective mass of the resonator modes.
 We find $m_{\mathrm{eff,S}}=3\,\mathrm{ng}$ and $m_{\mathrm{eff,A}}=4\,\mathrm{ng}$, for a defect separation of 1 unit cell in the direction  $\theta=13.9\deg$.
 Details on the procedure are presented in Appendix \ref{app:meff}.
For comparison, Table \ref{tab:table-1} summarizes simulated values of the effective mass for the different geometries.
These are extracted from the normalized displacement profile $\phi_i$ using $ m_{\mathrm{eff},i} \equiv \rho \langle \phi_i^2 \rangle$, where $\rho\approx 3200~\mathrm{kg/m^3}$ is the material's mass density.
The simulated values for the hybrid modes are of the order $2\,\mathrm{ng}$, close to twice the simulated effective mass of a single defect resonator, 0.92~ng.
Notably, as the defect separation becomes smaller, the mass of the true hybrid eigenmodes begins to deviate from this relation, indicating stronger mixing of the mode shapes.
We attribute the discrepancy with the (larger) measured masses to an imperfect overlap between the optical beam and the mechanical mode in the measurement (see Appendix \ref{app:lumpedElement}).

\begin{table}
\begin{tabular}{c|cc|cc|cc|cc}
$\theta$ & \multicolumn{2}{c}{0 unit cells} & \multicolumn{2}{c}{1 unit cell} & \multicolumn{2}{c}{2 unit cells} & \multicolumn{2}{c}{3 unit cells}\\
\hline
& $m_{\mathrm{eff,S}}$ & $m_{\mathrm{eff,A}}$ & $m_{\mathrm{eff,S}}$ & $m_{\mathrm{eff,A}}$ &$m_{\mathrm{eff,S}}$ & $m_{\mathrm{eff,A}}$ &$m_{\mathrm{eff,S}}$ & $m_{\mathrm{eff,A}}$\\
$0 \deg $& 2.6 & 1.8 & 1.9 & 1.7 & 1.7 & 1.9 & 1.9 & 1.8\\
$14 \deg$ &--&--&1.7&2.0&--&--&1.8&1.9\\
$30 \deg$ &--&--&1.7&2.0&1.8&1.8&1.9&1.8\\
$46 \deg$ &--&--& 1.7&2.0&--&--&1.8&1.9\\
\end{tabular}
\caption{\label{tab:table-1} Simulated effective masses (in nanograms) for the studied membrane devices. We assume nitride thicknesses of $14~\mathrm{nm}$ and a material density of $\rho=3200~\mathrm{kg/m^3}$.}
\end{table}

Piecing together the linear properties of our devices, we can estimate the force noise of power spectral density, $S_\mathrm{FF}$, of our devices. For a phononic dimer with $\theta=13.9\deg$ orientation and one unit cell separation, the force noise at room temperature for the symmetric and antisymmetric modes is estimated to $S_\mathrm{FF_S}=(28.2\,\mathrm{aN)^2/Hz}$ and $S_\mathrm{FF_A}=(29.9\,\mathrm{aN)^2/Hz}$, while the projected force noise at $4\,\mathrm{K}$ is $S_\mathrm{FF_S}=(1.9\,\mathrm{aN)^2/Hz}$ and $S_\mathrm{FF_A}=(2.0\,\mathrm{aN)^2/Hz}$.
We can furthermore evaluate the $Qf$-product, and find values in excess of $10^{14}~\mathrm{Hz}$. 
At room temperature, this corresponds to more than 15 coherent oscillations, or $10^3$ oscillations at 4~K.

\section{Geometric nonlinearities}
\label{sec:nonlinearities}

Some of the applications of interest -- parametric spin detection \cite{Kosata_2020} in particular -- require large-amplitude oscillations of a resonator mode.
We therefore proceed to characterizing the mechanical nonlinearities of the system.
High-aspect ratio string and drum resonators are known to exhibit a Duffing nonlinearity \cite{Schmid2011,Fong_2012, Hocke_2014}.
It originates from the modes' geometric nonlinearity, that elongates the device by an amount proportional to the {square} of the modes' excitation amplitude.
This leads to a nonlinear contribution to the elongation energy \cite{Ugural2017}
\begin{equation}
\label{eq:nonlinenergy}
    E_\mathrm{nlin}= \frac{E}{8(1-\nu^2)}
        \left \langle 
            \left[ 
                \left(\frac{\partial w}{\partial x}\right)^2 +
                \left(\frac{\partial w}{\partial y}\right)^2 
            \right]^2
            \right\rangle.
\end{equation}
We furthermore introduce a phenomenological nonlinear damping parameter $\gamma_i^\mathrm{snl}$, and write the equation of motion for the  amplitude of mode $i\in\{A,S\}$ as
\begin{equation}\label{eq:nl1}
    \ddot{u}_i+\left(\Gamma_i+\gamma_i^\mathrm{snl}u_i^2\right)\dot{u}_i+\Omega_i(\Omega_i+2\omega_i^\mathrm{sD}u_i^2)u_i=0.
\end{equation}
The ``self''-Duffing parameter $\omega_i^\mathrm{sD}$ directly indicates the frequency shift per quadratic displacement, and is linked to the standard Duffing parameter $\alpha$ by the relation
\begin{align}
\label{eq:selfDuffingDefinition}
\alpha=2m_{\mathrm{eff},i}\Omega_i\omega_i^\mathrm{sD}.
\end{align}
We note that the nonlinear parameters $\omega_s^\mathrm{sD}$ and $\gamma_i^\mathrm{snl}$ of the hybrid modes can also be connected to the nonlinearities of the local modes \cite{Kosata_2020}.

The nonlinearities  introduce an amplitude dependence in both the damping and the resonance frequency. 
In the present case, we can show (cf. Appendix \ref{app:solem}) that a free mechanical ringdown starting from a large initial amplitude $A_{i,0}$ follows $u_i(t)=A_i(t)\cos(\Omega_i(t)\cdot t)$, with 
\begin{equation}\label{eq:nl2}
    A_i(t)=\frac{A_{i,0}e^{-\frac{\Gamma_i}{2}t}}{\sqrt{1+\frac{\gamma_i^\mathrm{snl}}{4\Gamma_i}A_{i,0}^2\left(1-e^{-\Gamma_i t}\right)}}.
\end{equation}
and 
\begin{equation}\label{eq:nl3}
    \Omega_i(t) = \Omega_i+\underbrace{\frac{3}{4}\omega_i^\mathrm{sD}A_i^2(t)}_{\delta\Omega_i(A_i)},
\end{equation}
where the latter corresponds to the backbone curve of a Duffing oscillator \cite{Polunin_2016}.

To extract the nonlinear parameters $\omega^\mathrm{sD}_i$ and $\gamma_i^\mathrm{snl}$, we again drive the mechanical modes and then let them decay freely.
We monitor  the time-evolution of both amplitude and phase (and thereby instantaneous frequency). 
Since the nonlinear effects depend on the oscillation amplitude, we implement an absolute calibration of the interferometer signal in units of mechanical displacement (see Appendix \ref{app:calibration}).

\begin{figure}[htb]
\center
\includegraphics[scale=1]{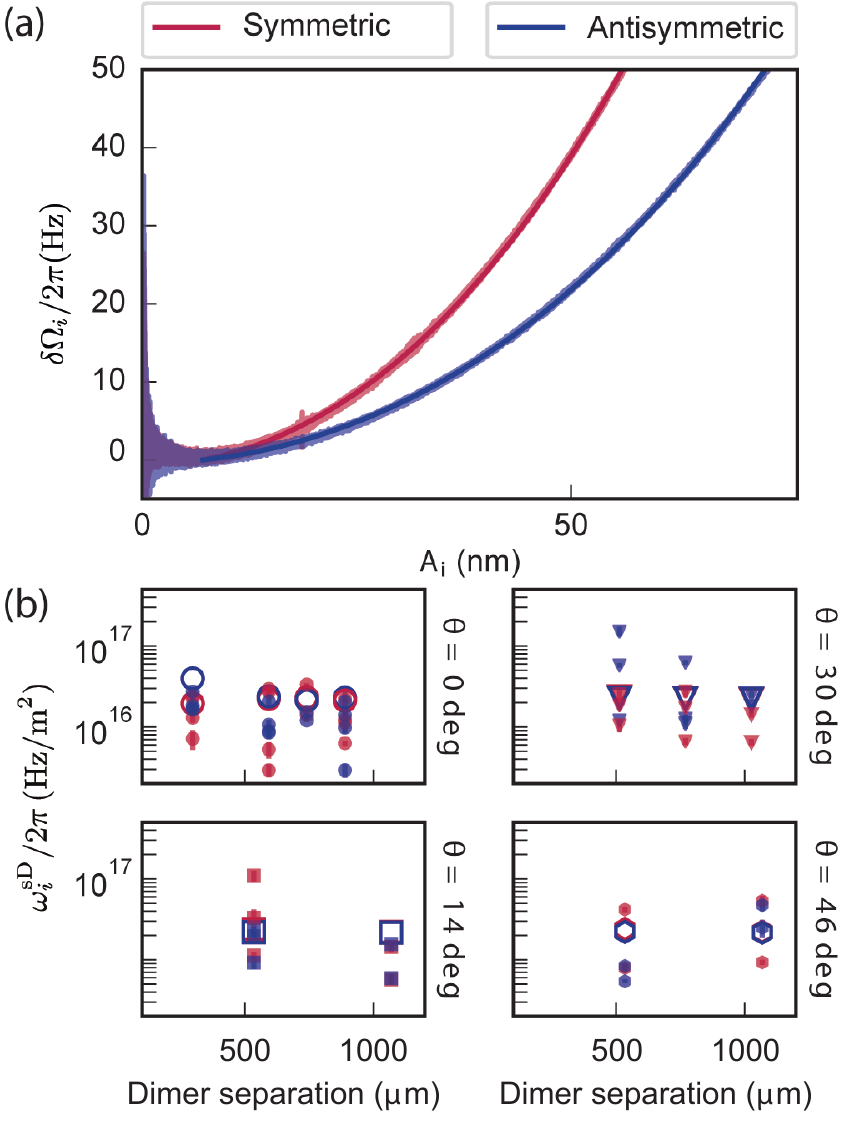}
\caption{Duffing parameter measurements and simulations. (a) Frequency shift as a function of the amplitude ($A_i$) for the geometry $\theta\approx14\,\deg$, $d_{\mathrm{sep}}=1\mathrm{\,unit\,cell}$. The frequency shift is extracted from the time derivative of the phase and is plotted against the fit of the amplitude decay. The light curves are the data, the darker lines the fit. (b) Duffing parameters for the symmetric ($\omega_S^\mathrm{sD}$), red points, and antisymmetric ($\omega_A^\mathrm{sD}$), blue points, mode for all the geometries. Each relative angle ($\theta$) is associated with a symbol, the dimer separation ($d_\mathrm{sep}$) is reported on the x axis.}
\label{f:fig3}
\end{figure}

\begin{figure}[htb]
\center
\includegraphics[scale=1]{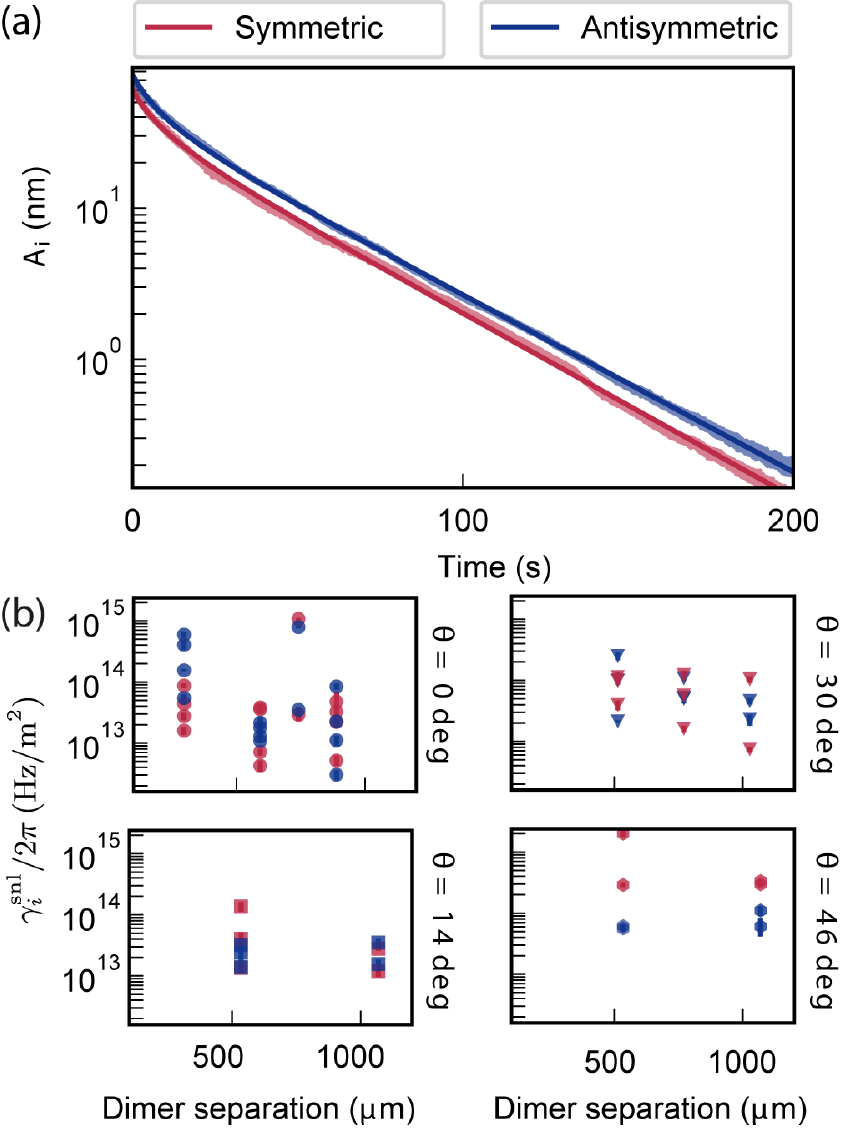}
\caption{Nonlinear damping measurements. (a) Amplitude decay $A_i(t)$ for a device of ${\theta=14^\circ}$ orientation and ${d_\mathrm{sep}=1\,\mathrm{unit\,cell}}$. The red and blue light lines are the amplitudes as a function of time of the symmetric and antisymmetric mode, respectively. The darker lines are the fitting curves. (b) Nonlinear damping values for the symmetric ($\gamma_\mathrm{S}^\mathrm{snl}$), red points, and antisymmetric ($\gamma_\mathrm{A}^\mathrm{snl}$), blue points, mode for all the geometries. Each relative angle ($\theta$) is associated with a symbol, the dimer separation ($d_\mathrm{sep}$) is reported on the x axis.}
\label{f:fig4}
\end{figure}

In our heterodyne detection scheme the interferometer arm length difference is not locked.
Fluctuations in phase due to change in the arms length are imprinted on both the phase of the carrier beat note and of the mechanical sidebands.
We eliminate such fluctuations in post processing, by referring the phase of the mechanical mode to the phase of the carrier beat note.
To extract the instantaneous mechanical resonance frequency $\Omega_{i}(t)$, we compute the numerical derivative of the so-corrected mechanical phase. 

From the time evolution of both frequency and amplitude, we can reconstruct the frequency shift as a function of the membrane oscillation amplitude, as shown in Fig.~\ref{f:fig3}-a.
We find excellent agreement of the data with a quadratic frequency shift as prescribed by eq.~(\ref{eq:nl3}), for oscillation amplitudes up to $\sim 85\mathrm{nm}$.
To fit the data to even larger amplitudes we introduce a phenomenological term proportional to $A_i^4(t)$.
Figure~\ref{f:fig3}-b summarizes the Duffing parameter obtained for all investigated geometries.

It is instructive to compare the Duffing parameters of phononic dimer structures to those of standard silicon nitride mechanical resonators \cite{Unterreithmeier2009a}.
The Duffing parameter associated with the fundamental mode of a tensioned square membrane is  ${\alpha \approx 5 E m_\mathrm{eff}\pi^4/(12\rho L^4)}$, where $L$ is the side length of the membrane (Appendix F). 
Using the definition of the self-Duffing parameter (cf. eq. (\ref{eq:selfDuffingDefinition})), and assuming a resonator length of $200~\mathrm{\mu m}$, $\bar{\sigma}=1.27~\mathrm{GPa}$ in-plane stress, $270~\mathrm{GPa}$ Young's modulus and $3200~\mathrm{kg/m^3}$ density, we find that $\omega_\mathrm{square}^\mathrm{sD}/2\pi\approx 10^{16}~\mathrm{Hz/m^2}$, comparable to the values of self-Duffing parameters measured for our devices.

Additionally, we simulate the Duffing parameters for the symmetric and antisymmetric modes.
Instead of a transient analysis \cite{Putnik_2016}, we are using a simple perturbative approach, in which we evaluate the nonlinear energy of eq.~(\ref{eq:nonlinenergy}) based on mode shapes obtained disregarding this term (see Appendix \ref{app:Duffingsimulation} for details).
We use an analytic model for the Duffing parameter of a tensioned doubly-clamped beam \cite{Hocke_2014} to validate our simulations, finding agreement between analytic and simulated values to within a few percent.
As shown in Fig.~\ref{f:fig3}-b, the predicted Duffing values for our much more complex geometry agree with our measurements within the experimental scatter.
As shown in Fig.~\ref{f:fig4}-a, the ringdowns also clearly show a non-exponential decay of the oscillation with time for amplitudes exceeding some $10~\mathrm{nm}$.
The behaviour is excellently reproduced using the model of eq.~(\ref{eq:nl2}).
We can extract the nonlinear damping parameter for all geometries for both the symmetric and antisymmetric mode, as shown in Fig.~\ref{f:fig4}-b.

Finally, we study cross-nonlinearities, i.~e.~the change in frequency and damping of one mode (index $i$), when the other mode (index $j$) is driven to large-amplitude oscillations.
Such effects have been observed in a variety of mechanical multi-mode devices, see e.~g.~refs.~\cite{Antoni_2012, Midtvedt_2014, Karabalin_2009, Westra_2010}.
In our case, they can again arise from the nonlinear elongation energy of eq.~(\ref{eq:nonlinenergy}) \cite{Midtvedt_2014}, but can also be obtained by transforming the standard (self-)nonlinear terms of the individual defect modes to the coupled basis \cite{Kosata_2020}.
The equation of motion can be written as 
\begin{equation}
\label{eq:nl4}
    \ddot{u}_i+(\Gamma_i+\gamma_{ij}^\mathrm{xnl}u_j^2)\dot{u}_i+\Omega_i\left(\Omega_i+2\omega_{ij}^\mathrm{xD}u_j^2\right)u_i=0
\end{equation}
where we refer to $\gamma_{ij}^\mathrm{xnl}$ and $\omega_{ij}^\mathrm{xD}$ as cross-nonlinear damping and cross-Duffing parameter, respectively.

\begin{figure}[htb]
    \center
    \includegraphics[scale=1]{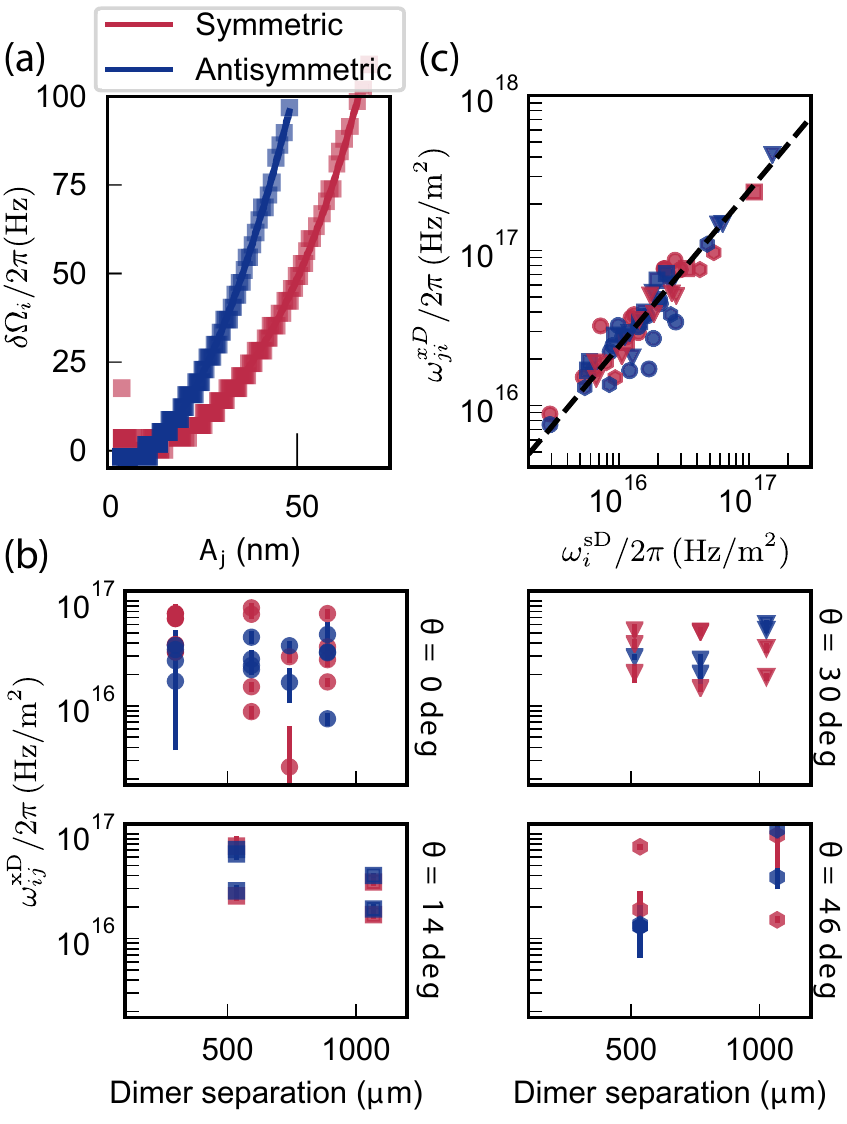}
    \caption{Cross-Duffing parameter measurements.
    (a) Frequency shift of the undriven mode as a function of the amplitude of the driven (red--symmetric, blue--antisymmetric) mode.
    Squares are data, lines quadratic fits, revealing the cross-Duffing parameter $\omega_{ij}^\mathrm{xD}$.
    (b) Cross-Duffing parameter for all measured geometries. 
    (c) Relation between self- and cross-Duffing parameters, where different symbols represent different angles as in panel b), with linear fit (dashed line).}
\label{f:fig6}
\end{figure}

As in the self-nonlinearities' case, the nonlinear terms introduce an amplitude dependence in the mechanical frequency and in the damping rate. 
From the solution of the equation of motion (see Appendix \ref{app:solem}) the amplitude-dependent damping rate is given by 
\begin{equation}\label{eq:nl5}
    \Gamma_i'=\Gamma_i+\frac{\gamma_{ij}^\mathrm{nl}}{2}A_j^2,
\end{equation}
{while the frequency shift is}
\begin{equation}\label{eq:nl6}
    \Omega_i'=\Omega_i+\underbrace{\frac{1}{2}\omega_{ij}^\mathrm{xD}A_j^2}_{\delta\Omega_i(A_j)}.
\end{equation}
To measure the effect of the cross-Duffing term in our structures, we analyze the thermal motion of the mode $i$ during the (nonlinear) ringdown of the mode $j$.
We extract the instantaneous frequency $\Omega'_i(t)$ by monitoring the thermal motion of mode $i$, and performing a Fourier transform of this time-trace every $0.28~\mathrm{s}$.
The frequency corresponding to the maximum of the Lorentzian gives the mechanical frequency during the considered time interval.
Relating the obtained frequency $\Omega'_i(t)$ with the mean value of the amplitude $A_j(t)$ during the same time interval, we can reconstruct the frequency shift as a function of the driven mode displacement amplitude (Fig. \ref{f:fig6}-a). 
From a quadratic fit we estimate the cross-Duffing parameter. 
All measured cross-Duffing parameters are shown in Fig.~\ref{f:fig6}-b. 

It is interesting to compare the strengths of the self-Duffing and the cross-Duffing effects.
For such a comparison, we plot the measured nonlinear parameters $\omega_{ji}^\mathrm{xD}$ and $\omega_i^\mathrm{sD}$ against each other, for the various characterized geometries (Fig.~\ref{f:fig6}-c).
We find a  ratio of  ${\omega_{ji}^\mathrm{xD}/\omega_i^\mathrm{sD} =2.6 \pm 0.5}$, compatible with a factor of  3 expected from theoretical analysis \cite{Kosata_2020}. The error corresponds to the $95\%$ confidence interval of the best fit.

To measure the cross-nonlinear damping, we again drive mode $j$ to large, now constant, amplitude $A_j$.
We then drive mode $i$ weakly, and let it decay, monitoring its ringdown, from which we extract the modified damping $\Gamma'_i$.
Fitting the obtained damping of the symmetric mode $\Gamma'_\mathrm{S}$ for different excitation amplitudes $A_\mathrm{A}$ of the antisymmetric mode using eq.~(\ref{eq:nl5}), we extract a cross nonlinear damping parameter $\gamma_{ij}^\mathrm{xnl}/2\pi=\,1\times 10^{13}\,\mathrm{Hz/m^2}$ for a device with $\theta=13.9\deg$ and a defect separation of 1 unit cell.

\begin{figure}[htb]
    \label{f:fig7}
    \center
    \includegraphics[scale=1]{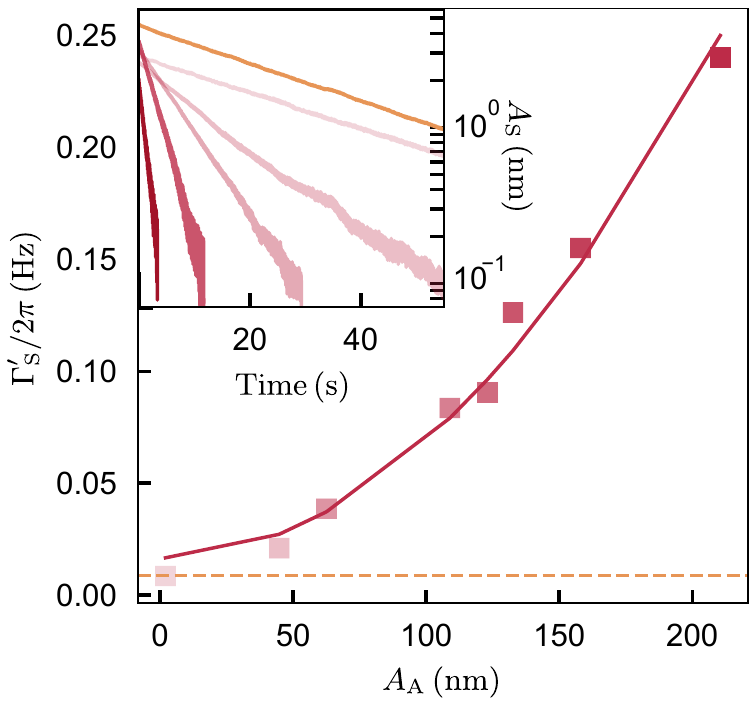}
    \caption{Cross-nonlinear damping measurements.  Linear damping estimated from the fit of the curve in the figure on the left, as a function of the strongly driven mode amplitude ($A_\mathrm{A}$). From the fit we estimate the cross-nonlinear damping, $\gamma_{SA}^\mathrm{xnl}/2\pi=5\times 10^{12}\,\mathrm{Hz/m^2}$. Yellow dashed line is the linear damping rate without any cross nonlinearities. Inset: Ringdown of the weakly driven mode ($A_\mathrm{S}(t)$) for several amplitude of the strongly driven one ($A_\mathrm{A}$). The strong drive is kept on during all the ringdown time. The yellow line is the linear ringdown in absence of a strong drive on the coupled mode.}
\end{figure}

\section{Conclusion and outlook}

In conclusion, we have developed and characterized phononic dimers consisting of two identical defects in a phononic crystal.
Their coupling leads to new  eigenmodes delocalized over both defects, with a frequency splitting that is tunable by the device geometry over the wide range $\sim 2-100~\mathrm{kHz}$. 
Importantly, the characteristic coherence of the soft-clamped mechanical modes is preserved. 
Investigations of the mechanical dynamics in the nonlinear regime have revealed a Duffing shift, as well as an amplitude-dependent damping rate.
As we show, these effects appear also on undriven modes, if another mode is excited to large amplitude.

The device geometry, tunability of the coupling, and long coherence time make this a promising platform for a number of applications. 
With a frequency splitting that easily exceeds the mechanical decoherence rate $k_\mathrm{B}T/\hbar Q$, multi-mode quantum optomechanics experiments \cite{Nielsen_2016} are conceivable, such as entanglement of mechanical resonators by reservoir engineering \cite{Ockeloen-Korppi_2017}. 
For the same purpose, the intrinsic nonlinearity of the mechanical system \cite{Mahboob_2014, Patil_2015} constitutes a resource yet untapped in the quantum regime.

Even when addressing only a single mechanical mode, the device geometry is conducive for  mechanically mediated transduction.
This includes electro-opto-mechanical transducers \cite{Midolo_2018} as well as optically detected force sensors.   
MRFM is an example for the latter, with particularly demanding requirements. 
In this case, the mechanical multi-mode structure enables the use of powerful parametric sensing protocols, given that the mechanical frequency splitting can be matched to the spin inversion frequency. Recent theoretical work by J.~Ko\v{s}ata \textit{et al.} \cite{Kosata_2020} evaluates the performance of the devices introduced here for parametric spin sensing.

\begin{acknowledgements}
The authors acknowledge discussions with J.~Ko\v{s}ata and A.~Eichler of ETH Zurich, and the wider collaboration within the SNF Sinergia project ``Zeptonewton force sensing on a membrane resonator platform''. 
We are grateful for assistance in device imaging by A.~Barg, D.~Mason and M.~Rossi, as well as discussions with Peter Degenfeld-Schonburg of Bosch GmbH. This work has received funding from the European Union's Horizon 2020 research and innovation programme (European Research Council (ERC) project Q-CEOM, grant agreement no.~638765, project  ULTRAFORS, grant agreement no.~825797, and FET proactive project HOT, grant agreement no.~732894), as well as the Swiss National Science Foundation under grant number 177198.
\end{acknowledgements}

\appendix
\section*{Appendices}
\section{Mapping to a lumped-element model}
\label{app:lumpedElement}
The dynamics of mechanical resonators with uniform in-plane tension in the linear regime is governed by the elastic equations of motion for a thin, isotropic rectangular plate \cite{Leissa1969} 
\begin{align}
\label{eq:fullisotropicEqMotion}
\bar{\sigma} \nabla^2 w=\rho \ddot w,
\end{align}
where $\bar{\sigma}$ is the in-plane tension and $\rho$ is the material density.
The total displacement field, $w$, can be described by a sum of normalized mode shapes, $\phi_n(x,y)$, and the associated out-of-plane amplitudes, $u_n$
\begin{align}
{w(x,y, t)=\sum_m u_m(t) \phi_m(x,y)}.
\end{align}
By convention, the mode functions are normalized such that $\max_V \phi(x,y)=1$, and thus $u_m(t)$ is the maximum out-of-plane displacement of the membrane at time $t$. 

Since we are primarily interested in the dynamics of a select few vibrational modes of the resonator, we employ a single-mode Galerkin discretization method \cite{NayfehPai2004}, which allows us to describe the vibrational modes in terms of effective resonator parameters. Multiplication with a test modefunction, $\phi_n(x,y)$, and subsequent integration over the resonator volume, yields
\begin{align}
\bar{\sigma}\sum_m u_m \int_V \phi_n \nabla^2 \phi_m~dV= \rho \sum_m \ddot u_m \int_V \phi_n \phi_m~dV.
\end{align}
Here we neglect the spatial dependence of the in-plane tension in our perforated devices, as we are primarily interested in  qualitative features of dimerization.

Using Green's first identity on the left-hand side of the equation, in conjunction with the boundary conditions for a plate clamped along all four edges (i.e., $w(x,y)|_\mathrm{all~edges}=0$), results in the following equation
\begin{align}
\label{eq:discreticedEOM1}
\sum_m u_m \bar{\sigma}\int_V \nabla\phi_n\cdot\nabla \phi_m~dV= \sum_m \ddot u_m \rho \int_V \phi_n \phi_m~dV.
\end{align}
We now define the effective spring and mass matrices
\begin{align}
K_{nm}:=\bar{\sigma}\langle \nabla\phi_n | \nabla\phi_m \rangle \\
M_{nm}:=\rho \langle \phi_n | \phi_m \rangle,
\end{align}
where $\langle\cdot\rangle$ denotes integration over the volume of the resonator. 
With these definitions, eq.~(\ref{eq:discreticedEOM1}) translates into the eigenvalue problem 
\begin{align}
Ku = \Omega^2 Mu,
\label{eq:eigenvalueEOM}
\end{align}
where $u$ is now a vector of mode amplitudes $u_m$.
The eigenvectors represent the solutions of eq.~(\ref{eq:fullisotropicEqMotion}), expanded in the test functions $\{\phi_m\}$, with an eigenfrequency $\Omega$.\\

\begin{figure}[h!]
\centering
\includegraphics[width=1\linewidth]{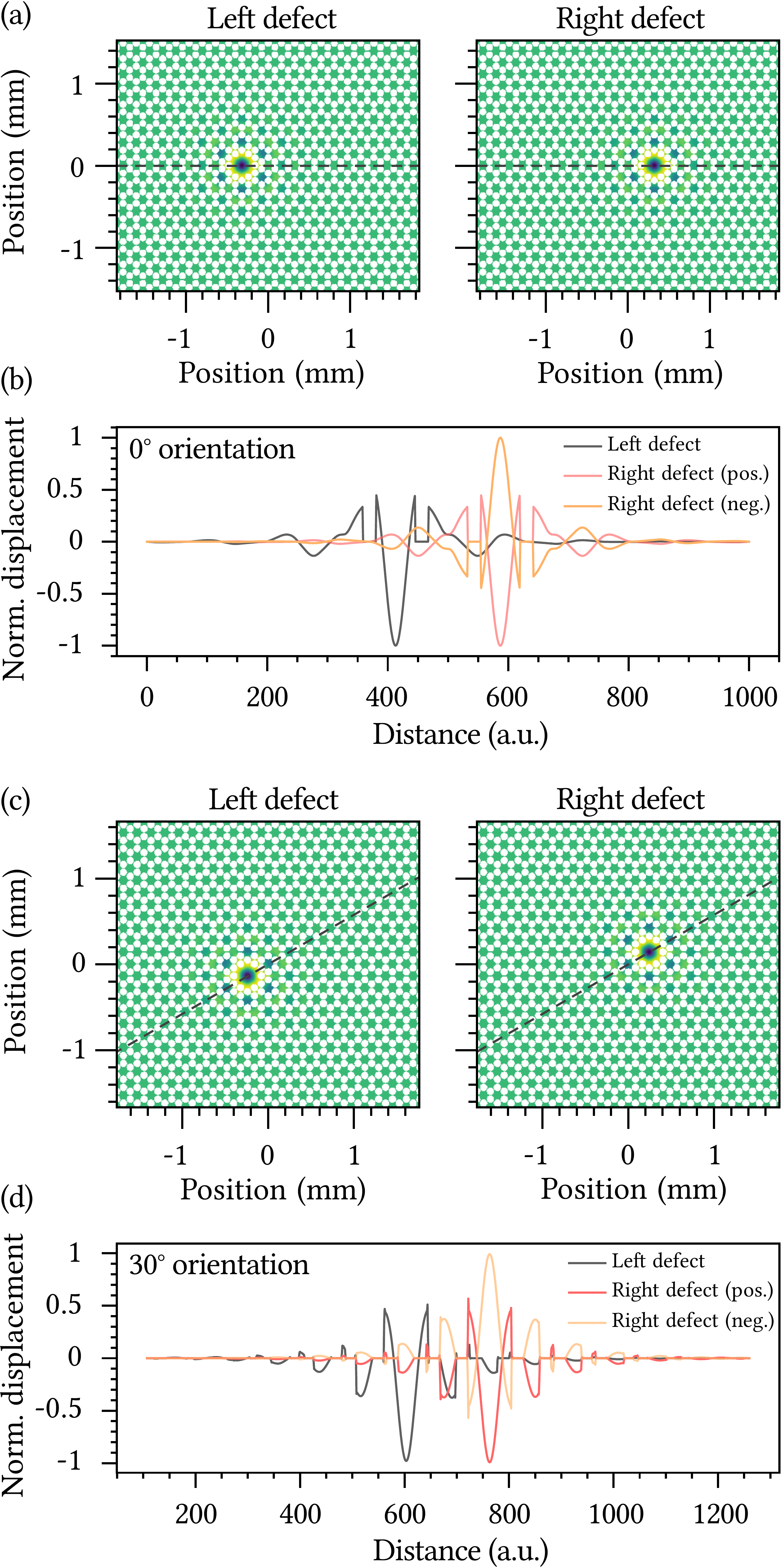}
\caption{Simulated displacement profiles of the individual defects associated with phononic dimers with orientations (a-b) $\theta=0\deg$ and (c-d) $\theta=30\deg$, and approximately one unit cell separation. A cross-sectional view of the (normalized) displacement fields for the defects to the left and right of the origin are shown in (b) and (d), along the dashed lines shown in (a) and (c), respectively. To illustrate the overlap between the displacement fields of the two defects for each geometry, the inverse of the displacement field of the right defect is shown in (b) and (d). For (b) $\theta=0\deg$ the displacement profile of the left defect is largely ''in-phase" with the inverse displacement profile of the right defect, while the opposite is the case for (d) $\theta=30\deg$.}
\label{f:modaloverlaps}
\end{figure}

We now consider the case of two coupled defects of a tensioned resonator. Following eq. (\ref{eq:eigenvalueEOM}) we arrive at the following set of equations
\begin{align}
K_{11}u_1 + K_{12}u_2 &= \Omega^2(M_{11}u_1+M_{12}u_2) \\
K_{21}u_1 + K_{22}u_2 &= \Omega^2(M_{21}u_1+M_{22}u_2).
\end{align}
Assuming identical defects, the diagonal (as well as off-diagonal) elements of the effective spring and mass matrices are identical (e.g. $K_{11}=K_{22}$, $K_{12}=K_{21}$, etc.). Taking the sum (difference) of these equations yields the resonance frequency for the symmetric (antisymmetric) mode
\begin{align}
\Omega_\mathrm{S} &= \sqrt{\frac{K_{11}+K_{12}}{M_{11}+M_{12}}} \\
\Omega_\mathrm{A} &= \sqrt{\frac{K_{11}-K_{12}}{M_{11}-M_{12}}}.
\end{align}
Defining the bare eigenfrequency of the individual defects as $\Omega_0\equiv \sqrt{K_{11}/M_{11}}$ and expanding the difference frequency $\Omega_\mathrm{S}-\Omega_\mathrm{A}$ to lowest order in $K_{12}/K_{11}$ and $M_{12}/M_{11}$ (corresponding to weakly coupled defects), we arrive at the following simplified expression for the resonance frequency splitting of the hybridized modes
\begin{align}
\frac{\Omega_\mathrm{A}-\Omega_\mathrm{S}}{\Omega_0}\approx\left(\frac{M_{12}}{M_{11}}-\frac{K_{12}}{K_{11}}\right).
\end{align}
The off-diagonal elements of the mass matrix corresponds to the overlap integral of the displacement fields associated with each defect. Evaluating the above equation for the four device geometries associated with Fig.~\ref{f:fig2}a correctly predicts the change in the ordering of the symmetric and antisymmetric mode. In a complementary analysis, we consider the displacement profiles for two of these configurations, namely for dimers with $\theta=0\deg$ and $\theta=30\deg$ orientations. We simulate the displacement profiles for each defect individually (see Fig.~\ref{f:modaloverlaps}a and Fig.~\ref{f:modaloverlaps}c), and plot the normalized displacement profiles of each defect along the axis connecting the two defects. For $\theta=0\deg$ (see Fig.~\ref{f:modaloverlaps}b) the displacement profiles are largely in-phase for the antisymmetric combination of displacement fields, while the  opposite is the case for $\theta=30\deg$ (see Fig. \ref{f:modaloverlaps}d). As a result, the order of the resonance frequencies of the hybrid modes is opposite for the two configurations.

Turning our attention to the effective modal masses of the symmetric and antisymmetric modes, and assuming negligible spatial overlap between the modes of interest and all other modes of the resonator, the effective mass takes the following simple form
\begin{align}
\label{eq:effMassDefinition}
m_{\mathrm{eff}, i} = \rho\int_V \phi_i^2~\mathrm{d}V,
\end{align}
where $i\in\left\{S,A\right\}$. The equation above is used in calculating the effective masses for all membrane devices (see Appendix \ref{app:simulation}).
We note that eq. (\ref{eq:effMassDefinition}) describes the effective mass of a vibrational mode, as seen by a point-like probe placed at the point of maximum out-of-plane displacement.
In our experiments, however, the membrane is probed by a Gaussian beam of finite dimensions, which is likely shifted from the point of maximum displacement.
This can be described by an overlap intergral between the displacement profile of the resonator and a Gaussian distribution, corresponding to the optical intensity profile. Hence, the measured effective mass can be expressed as follows

\begin{align}
\tilde{m}_{\mathrm{eff}, i} &= \rho\int \frac{(u_i \phi_i(x,y))^2}{\left(\int_A u_i \phi_i(x,y)\psi(x,y) \mathrm{d}A\right)^2}~\mathrm{d}V, \\
&= \left(\int_A \phi_i(x,y)\psi(x,y) \mathrm{d}A\right)^{-2}\rho\int \phi_i^2~\mathrm{d}V, \\
&=: \eta_{\mathrm{om}}^{-2} m_{\mathrm{eff}, i},
\end{align}
where $\psi(x,y)$ is the area-normalized Gaussian mode profile, and $\eta_{\mathrm{om}}$ is  defined as the overlap integral between the normalized optical and mechanical modes.
A similar argument can be made for the measured out-of-plane displacement.
As we will see in Appendix \ref{app:calibration}, this simple fact is of importance as it pertains to the amplitude calibration method employed in our study.
\section{Interferometric setup}
\label{app:setup}

\begin{figure}[htb]
\centering
\includegraphics[scale=1]{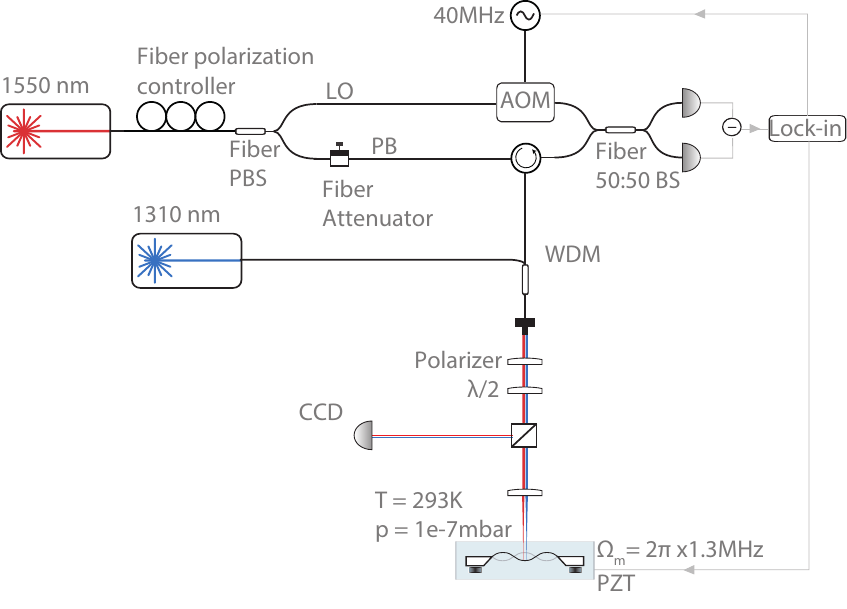}
\caption{Detailed experimental setup. LO, local oscillator; PB, probe beam; AOM, acoustooptic modulator; PBS, polarizing beam splitter; BS, beam splitter; WDM, wavelength division multiplexer; $\lambda/2$, half wave plate; CCD, imaging sensor; $T$, temperature; $p$, pressure; PZT, piezoelectric actuator; $\Omega_m$, mechanical frequency. Dark grey lines represent fiber connections, light grey lines electric connections, red and blue lines light free space path.}
\label{f:figA2}
\end{figure}

The setup employed for characterizing the resonators consists of a Mach-Zender interferometer, described in Fig. \ref{f:figA2}. The whole setup is fiber based. The light source is a fiber laser with 1550 nm wavelength. A fiber polarization controller with a polarizing beam splitter is used to define the amount of input light in the local oscillator  and in the probe beam arm. After the splitting, the probe beam is sent into the port 1 of a fiber circulator. The light leaves the circulator at port 2 of the circulator and is sent free-space through a fiber collimator towards the membrane, which is placed in a vacuum chamber at room temperature. A turbo pump is mounted  directly on the vacuum chamber producing vibrations of the setup. Such vibration appears as satellite peaks around the mechanical modes. The light goes through a polarizer and a half wave plate to correct the polarization and it is reflected back from the membrane. The reflected light goes back into the port 2 of the circulator and after the port 3 is recombined with the LO, previously shifted by $40\,\mathrm{MHz}$ using an acoustooptic modulator (AOM), on a balanced detector. The heterodyne detection photocurrent is input to a lock-in amplifier.

We drive the membrane by a piezoelectric actuator. Both the $40\,\mathrm{MHz}$ modulation of the AOM and the driving signal of the piezoelectric actuator are generated by the same lock-in amplifier.

A torch light is used to image the defect of the membrane on a CCD camera. Such an imaging is used to align the probe laser to the center of the defect. Since 1550 nm wavelenght is not detected by the CCD camera used, we use an additional $1310\,\mathrm{nm}$ diode laser for this purpose. Since the two beams are launched to free space from the same fiber and collimation optics, we assume that the $1310\,\mathrm{nm}$ beam spot postion, visible on the CCD camera, coincides with the $1550\,\mathrm{nm}$ beam spot position. We select the wavelength to send back into the interferometer using a wavelength division multiplexer.

\section{Calibration}\label{app:calibration}
Reflection off the membrane converts the oscillation of the membrane in a phase modulation of a probe beam. This phase modulation is converted into amplitude modulation of a photocurrent (amplified and measured as a voltage) by beating the probe beam with a local oscillator. To express the amplitude in displacement units (m) the data must be calibrated. We assume that, without applying any forces, the mechanical resonator is in thermal equilibrium with the temperature $T$ of the setup, i.e., the spectral area of a mechanical mode coincides with its thermal energy
\begin{equation}\label{cl_eq1}
\mathrm{Area}_\mathrm{th}\equiv\int S_{xx}^2\frac{d\Omega}{2\pi}=\frac{k_BT}{m_{\mathrm{eff},i}\Omega_i^2}
\end{equation}
where $k_B$ is the Boltzmann constant, $m_{\mathrm{eff},i}$ is the effective mass of the mechanical oscillator (coming from numerical simulations described in Appendix \ref{app:simulation}) and $\Omega_i$ is the mechanical frequency. Before starting a measurement, we save a thermal spectrum in Volt-RMS units ($S_{VV}^{\mathrm{th}}$). From its area a conversion factor ($g$) is derived
\begin{equation}\label{cl_eq2}
g^2\left[\frac{V^2}{m^2}\right]=\frac{m_{\mathrm{eff},i}\Omega_i^2}{k_BT}\int S_{VV}^\mathrm{th}\frac{d\Omega}{2\pi}.
\end{equation}
Once the calibration parameter $g$ is estimated from the thermal state of the mechanical oscillator, we can calibrate the data by simply
\begin{equation}\label{cl_eq3}
S_{xx}^\mathrm{meas}=g^{-2}S_\mathrm{VV}^\mathrm{meas}
\end{equation}
However, this calibration is correct only as long as the amount of reflected light doesn't change. We need a way to correct for fluctuations in light intensity during the measurement. To this end, we use the amplitude of the $40\,\mathrm{MHz}$ "carrier", i.e., the beat note between the local oscillator and the unmodulated light reflected off the membrane. The ratio between the peak value of the carrier during the initial thermal trace, and the peak value at any time $t$ during the ringdown, gives us a correction parameter to apply in our calibration.
\section{Effective mass measurement}\label{app:meff}
To calibrate our data (Appendix \ref{app:calibration}) we assume that the effective mass in the measurement coincides with the simulated one, $\tilde{m}_{\mathrm{eff},i}=m_{\mathrm{eff},i}$. It is instructive proceed through an alternative calibration process to estimate the effective mass from the calibrated spectrum and compare it with the simulated values.

The Brownian motion $\delta u_i$ of the membrane will be transduced in two mechanical sidebands at ${40\,\mathrm{MHz}\pm\Omega_i/2\pi}$ in the heterodyne signal. The amplitude of the mechanical sidebands is given by the first order Bessel function $J_1(\beta)$, where $\beta$ is the modulation depth. Since we are in the limit where the Bessel function is linear, the measured spectrum $S_{VV}$ in radians is given by
\begin{equation}
    S_{\varphi\varphi}=\frac{S_{VV}}{\int S^2_{VV_\mathrm{mech}}\frac{d\Omega}{2\pi}}\frac{\beta^2}{2}
\end{equation}
where $\beta$ satisfies the relation
\begin{equation}
    \left|\frac{J_0(\beta)}{J_1(\beta)}\right|^2=\frac{\int S^2_{VV_\mathrm{car}}\frac{d\Omega}{2\pi}}{\int S^2_{VV_\mathrm{mech}}\frac{d\Omega}{2\pi}}.
\end{equation}
The numerator and denominator are the area under the $40\,\mathrm{MHz}$ carrier peak, and the mechanical mode, respectively.
If the mechanical mode and the optical beam are perfectly overlapped, we can assume that
\begin{equation}
    \delta\varphi=\frac{4\pi}{\lambda}u_i
\end{equation}
where $\lambda$ is the laser wavelength. From this relation the displacement spectrum in $\mathrm{m}$ is obtained as
\begin{equation}
    S_{xx}=S_{\varphi\varphi}\left(\frac{\lambda}{4\pi}\right)^2.
\end{equation}
The detected motion is the Brownian motion of the mechanical oscillator at room temperature. The area under the mechanical mode expressed in $\mathrm{m}^2$ must be equal to the thermal energy of the oscillator, see eq. (\ref{cl_eq1}). From this relation we can estimate the effective mass as
\begin{equation}
    m_{\mathrm{eff},i}=\frac{k_\mathrm{B}T}{\mathrm{Area}_\mathrm{mech}\Omega_i^2}.
\end{equation}
Leaving the mode in its thermal equilibrium, we save a thermal trace of both the mechanical mode and the carrier for 15 minutes. We divide the data in shorter traces of 1 minute length and perform the FFT. For each of them we extract the effective mass as described above and we average all obtained values. 
The effective masses measured for the membrane with $14\deg$ relative angle, 1 unit cell separation are $m_S=3\,\mathrm{ng}$ and $m_A=4\,\mathrm{ng}$. Those values are larger than the simulated one, even though the order of magnitude is the same. We attribute this deviation to an imperfect overlap between the optical beam and the mechanical mode.
\section{Solution of the nonlinear equation of motion}\label{app:solem}
We study the nonlinear dynamics of a mechanical oscillator considering first only the self-nonlinearities, then only the cross-nonlinearities.
All the functions used to extract the nonlinear parameters come from the solution of the equation of motion in the mentioned cases.

Here we start solving the equation of motion considering only the self-nonlinearities:
\begin{equation}\label{MS_eq1}
\ddot{u_i}+(\Gamma_i+\gamma_i^\mathrm{nl}u_i^2)\dot{u_i}+\Omega_i(\Omega_i+2\omega_i^\mathrm{sD}u_i^2)u_i = 0.
\end{equation}
We assume the oscillator initially displaced from its equilibrium position
\begin{equation}\label{MS_eq2}
u_i(0)=A_{i,0}.
\end{equation}
To solve the equation we apply the multiple scales method \cite{Nayfeh_perturbation}. This method is based on the distinction of two time scales, a fast one $T_0$, associated with the mechanical oscillation, and a slow one $T_1$, associated with both the dampings and the Duffing parameter.  Assuming that the dampings and the Duffing are small contributions, we write the equation highlighting the small terms with an $\epsilon$
\begin{equation}\label{MS_eq3}
\ddot{u}_i+\epsilon\tilde{\Gamma}_i\dot{u}_i+\epsilon\tilde{\gamma}_i^\mathrm{nl}u_i^2\dot{u}_i+\Omega_i^2u_i+2\epsilon\Omega_i\tilde{\omega}_i^\mathrm{sD}u_i^3=0,
\end{equation}
with $\epsilon\tilde{\Gamma}_i=\Gamma_i$, $\epsilon\tilde{\gamma}_i^\mathrm{nl}=\gamma_i^\mathrm{nl}$ and $\epsilon\tilde{\omega}_i^\mathrm{sD}=\omega_i^\mathrm{sD}$.
The solution will be a function of the new time scales
\begin{equation}\label{MS_eq4}
u_i(t)=u_i(T_0, \epsilon T_1), 
\end{equation}
where $T_0=t$ and $T_1=\epsilon t$. Using the chain rule yields
\begin{align}\label{MS_eq5}
\frac{d}{dt}&=\frac{\partial}{\partial T_0}+\epsilon\frac{\partial}{\partial T_1},\\
\label{MS_eq6}
\frac{d^2}{dt^2}&=\frac{\partial^2}{\partial T_0^2}+2\epsilon\frac{\partial^2}{\partial T_0\partial T_1}.
\end{align}
Substituting eq. (\ref{MS_eq5}, \ref{MS_eq6}) in eq. (\ref{MS_eq3}) and introducing the uniform approximate solution
\begin{equation}\label{MS_eq8}
u_i(T_0,\epsilon T_1)=u_{i,0}(T_0,T_1)+\epsilon u_{i,1}(T_0,T_1),
\end{equation}
we can write the equation of motion as
\begin{widetext}
\begin{equation}\label{MS_eq9}
    \frac{\partial^2 u_{i,0}}{\partial T_0^2}+\Omega_i^2u_{i,0}+\epsilon\Big[\frac{\partial^2u_{i,1}}{\partial T_0^2}+2\frac{\partial^2u_{i,0}}{\partial T_0\partial T_1}+\tilde{\Gamma}_i\frac{\partial u_{i,0}}{\partial T_0}+\tilde{\gamma}_i^\mathrm{nl}u_{i,0}^2\frac{\partial}{\partial T_0}u_{i,0}+\Omega_i^2u_{i,1}+2\Omega_i\tilde{\omega}_i^\mathrm{sD}u_{i,0}^3\Big]=0.
\end{equation}
\end{widetext}
Finding the solution corresponds to solving the system
\begin{equation}\label{MS_eq10}
\begin{cases}\
    \frac{\partial^2u_{i,0}}{\partial T_0^2}+\Omega_i^2u_{i,0}=0,\\
    \begin{aligned}
    \frac{\partial^2u_{i,1}}{\partial T_0^2}+\Omega_i^2u_{i,1} = &-2\frac{\partial^2u_{i,0}}{\partial T_0\partial T_1}-\tilde{\Gamma}_i\frac{\partial u_{i,0}}{\partial T_0}+\\
    &-\tilde{\gamma}_i^\mathrm{nl}u_{i,0}^2\frac{\partial}{\partial T_0}u_{i,0}-2\Omega_i\tilde{\omega}_i^\mathrm{sD}u_{i,0}^3.
    \end{aligned}
\end{cases}
\end{equation}
The solution of eq. (\ref{MS_eq10}-1) is simply
\begin{equation}\label{MS_eq11}
    u_{i,0} =A_i(T_1)\cos(\Omega_iT_0+\beta_i(T_1)).
\end{equation}
We can substitute eq. (\ref{MS_eq11}) in eq. (\ref{MS_eq10}-2), which becomes
\begin{align}\label{MS_eq13}
    \frac{\partial^2u_{i,1}}{\partial T_0^2}+\Omega^2_iu_{i,1}& =\Omega_iF_1\sin(\Omega_iT_0+\beta_i(T_1))+\\
    &+\Omega_iF_2\cos(\Omega_iT_0+\beta_i(T_1))+\\
    &+\Omega_iF_3\sin(3\Omega_iT_0+3\beta_i(T_1))+\\
    &+\Omega_iF_4\cos(3\Omega_iT_0+3\beta_i(T_1)),
\end{align}
where

\begin{align}
F_1&=2\frac{\partial A_i(T_1)}{\partial T_1}+\tilde{\Gamma}_iA_i(T_1)+\frac{1}{4}\tilde{\gamma}_i^\mathrm{nl}A_i^3(T_1),\\
F_2&=2A_i(T_1)\frac{\partial\beta_i(T_1)}{\partial T_1}-\frac{3}{2}\tilde{\omega}_i^\mathrm{sD}A_i^3(T_1),\\
F_3&=\frac{1}{4}\tilde{\gamma}_i^\mathrm{nl}A_i^3(T_1),\\
F_4&=-\frac{1}{2}\tilde{\omega}_i^\mathrm{sD}A_i^3(T_1).
\end{align}
To avoid secular terms, we need to set the coefficients in front of $\sin(\Omega_iT_0+\beta_i(T_1))$ and $\cos(\Omega_iT_0+\beta_i(T_1))$ equal to zero:
\begin{equation}\label{MS_eq14}
\begin{cases}
    2\frac{\partial A_i(T_1)}{\partial T_1}+\tilde{\Gamma}_iA_i(T_1)+\frac{1}{4}\tilde{\gamma}_i^\mathrm{nl}A_i^3(T_1)= 0 \\
    2A_i(T_1)\frac{\partial\beta_i(T_1)}{\partial T_1}-\frac{3}{2}\tilde{\omega}_i^\mathrm{sD} A_i(T1)^3 = 0.\\
\end{cases}
\end{equation}
From the solution of eq. (\ref{MS_eq14}-1) we find the time evolution of the displacement amplitude
\begin{equation}\label{MS_eq16}
    A_i^2(T_1)=\frac{ce^{-\tilde{\Gamma}_iT_1}}{1-c\frac{\tilde{\gamma}_i^\mathrm{nl}}{4\tilde{\Gamma}_i}e^{-\tilde{\Gamma}_iT_1}},
\end{equation}
where $c$ will be determined applying the initial condition. Inserting eq. (\ref{MS_eq16}) in eq. (\ref{MS_eq14}-2) we find 
\begin{equation}\label{MS_eq18}
    \beta_i(T_1)=\frac{3}{4}\tilde{\omega}_i^\mathrm{sD}\frac{4}{\tilde{\gamma}_i^\mathrm{nl}}\log\left(1-c\frac{\tilde{\gamma}_i^\mathrm{nl}}{4\tilde{\Gamma}_i}e^{-\tilde{\Gamma}_iT_1}\right)+\Phi. 
\end{equation}
We are interested in the zero-th order approximation, which means that the solution is obtained by substituting the expression for $A_i(T_1)$ and $\beta_i(T_1)$ into $u_{i,0}$ and considering only the $\epsilon^0$ term. To satisy the initial condition we define the constant value $\Phi$ in such a way that it is equal to $\beta_i(0)$. In this way the initial condition is satisfied imposing $A_i(0)=A_{i,0}$.
This leads to the solution
\begin{widetext}
\begin{equation}\label{MS_eq21}
    u_i(t)=\underbrace{\frac{A_{i,0}e^{-\frac{\Gamma_i}{2}t}}{\sqrt{1+\frac{\gamma_i^\mathrm{nl}}{4\Gamma_i}A_{i,0}^2\left(1-e^{-\Gamma_i t}\right)}}}_{A_i(t)}\cos\underbrace{\left(\Omega_it+\frac{3}{4}\frac{4\omega_i^\mathrm{sD} }{\gamma_i^\mathrm{nl}}\log\left(1+\frac{\gamma_i^\mathrm{nl}}{4\Gamma_i}A_{i,0}^2\left(1-e^{-\Gamma_i t} \right)\right)+\Phi\right)}_{\varphi_i(t)},
\end{equation}
\end{widetext}
where $A_i(t)$ describes the nonlinear amplitude decay while the time derivative of $\varphi(t)$ describes the shift of the mechanical frequency as a function of the amplitude
\begin{equation}\label{MS_eq22}
\Omega_i'=\frac{d\varphi_i}{dt}=
\Omega_i+\frac{3}{4}\omega_i^\mathrm{sD}\underbrace{\frac{A_{i,0}^2e^{-\Gamma_i t}}{1+\frac{\gamma_i^\mathrm{nl}}{4\Gamma_i}A_{i,0}^2\left(1-e^{-\Gamma_i t}\right)}}_{A_i^2(t)}.
\end{equation}
Eq. (\ref{MS_eq22}) has the same form of the backbone equation. We should notice that the \textit{standard} backbone equation relates the maximum frequency shift with the maximum displacement amplitude of a driven oscillator. Extracting a backbone curve usually requires measuring the frequency response of the mechanical oscillator for several driving strengths. In our case, a single ringdown provides complete information about the frequency shift as a function of displacement amplitude. A similar approach is used in ref. \cite{Polunin_2016}.

\begin{figure}[htb]
\centering
\includegraphics[scale=1]{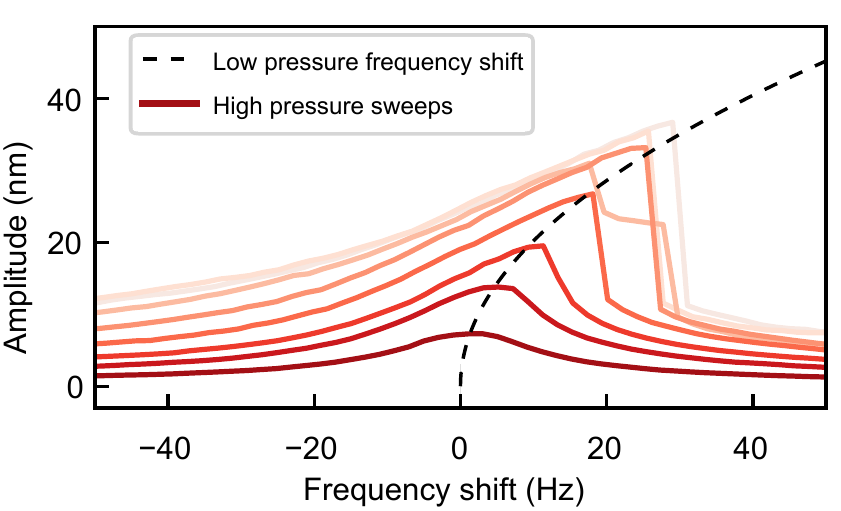}
    \caption{Frequency response of a driven Duffing oscillator at $3\times 10^{-3}\,\mathrm{mbar}$. For each color a driving force with a different amplitude was swept around the mechanical mode. Each curves is the mean value of 5 sweep curves performed with the same driving strength. The black dashed curve is the fit of the frequency shift as a function of amplitude obtained from the ringdown at low pressure for the same membrane.}
\label{f:figA1}
\end{figure}

In order to demonstrate the validity of the ringdown approach, we perform driven (swept) measurements, comparing the shark-fin like Duffing response to the backbone curve extracted from a ringdown.
To reduce the duration of the measurements, we increase the pressure in the vacuum chamber to $3\times 10^{-3}\,\mathrm{mbar}$. To extract such a curve the network analyser function of the lock-in amplifier was used. Then we compare the obtained curve with the frequency shift observed for a low pressure ringdown performed on the same membrane (Fig. \ref{f:figA1}).
As it arises from the data, the frequency shift during the ringdown follows the backbone curve.\\

Next we solve the equation of motion taking into account only the contribution of the cross-nonlinearities:
\begin{equation}\label{MS_eq23}
    \ddot{u}_i+(\Gamma_i+\gamma_{ij}^\mathrm{xnl}u_j^2)\dot{u}_i+\Omega_i(\Omega_i+2\omega_{ij}^\mathrm{xD}u_j^2)u_i=0.
\end{equation}
We impose the same initial condition and we assume that the external driven mode displacement is described by
\begin{equation}\label{MS_eq25}
    u_j=A_j\cos(\Omega_jt).
\end{equation}
As in the self-nonlinearities' case, we use the multiple scales method. At first, we highlight the small terms with an epsilon
\begin{equation}\label{MS_eq26}
    \ddot{u}_i+(\epsilon\tilde{\Gamma}_i+\epsilon\tilde{\gamma}_{ij}^\mathrm{xnl}u_j^2)\dot{u}_i+\Omega_i(\Omega_i+2\epsilon\tilde{\omega}_{ij}^\mathrm{xD}u_j^2)u_i=0.
\end{equation}
Following the same steps, we introduce the two time scales $T_0$ and $T_1$. Substituting the uniform approximate solution, eq. (\ref{MS_eq26}) becomes
\begin{widetext}
\begin{equation}
        \frac{\partial^2u_{i,0}}{\partial T_0^2}+\Omega_i^2u_{i,0}+\epsilon\Big[\frac{\partial^2u_{i,1}}{\partial T_0^2}+2\frac{\partial^2u_{i,0}}{\partial T_0 \partial T_1}+\tilde{\Gamma}_i\frac{\partial u_{i,0}}{\partial T_0}+\tilde{\gamma}_{ij}^\mathrm{xnl}A_j^2\cos^2(\Omega_jT_0)\frac{\partial u_{i,0}}{\partial T_0}+2\Omega_i\tilde{\omega}_{ij}^\mathrm{xD}A_j^2\cos^2(\Omega_jT_0)u_{i,0}+\Omega_i^2u_{i,1}\Big]=0.
\end{equation}
\end{widetext}
Solving this equation corresponds to solving the system
\begin{equation}\label{MS_eq29}
\begin{cases}
    \begin{aligned}
    \frac{\partial^2u_{i,0}}{\partial T_0^2}+\Omega_i^2u_{i,0}=&\,0\\
    \frac{\partial^2 u_{i,1}}{\partial T_0^2}+\Omega_i^2u_{i,1}=&-2\frac{\partial^2 u_{i,0}}{\partial T_0\partial T_1}-\tilde{\Gamma}_i\frac{\partial u_{i,0}}{\partial T_0}+\\
    &-\tilde{\gamma}_{ij}^\mathrm{xnl}Aj^2\cos^2(\Omega_jT_0)\frac{\partial u_{i,0}}{\partial T_0}+\\
    &-2\Omega_i\tilde{\omega}_{ij}^\mathrm{xD}Aj^2\cos^2(\Omega_jT_0)u_{i,0}.
    \end{aligned}
\end{cases}
\end{equation}
Inserting the solution of equation of eq. (\ref{MS_eq29}-1)
\[u_{i,0}=A_i(T_1)\cos(\Omega_iT_0+\beta_i(T_1)),\] 
into eq. (\ref{MS_eq29}-2)
\begin{equation}\label{MS_eq32}
    \begin{aligned}
    \frac{\partial^2u_{i,1}}{\partial T_0^2}+\Omega_i^2u_{i,1}&=F_1\sin(\Omega_iT_0+\beta_i(T_1))+\\
    &+F_2\cos(\Omega_iT_0+\beta_i(T_1)),
    \end{aligned}
\end{equation}
where
\[
\begin{split}
F_1&=2\Omega_i\frac{\partial A_i(T_1)}{\partial T_1}+\Omega_i\tilde{\Gamma}_iA_i(T_1)
+\tilde{\gamma}_{ij}^\mathrm{xnl}Aj^2A_i(T_1)\Big[\frac{1}{2}+\\
&+\cos(2\Omega_j)\Big]\\
F_2&=2\Omega_iA_i(T_1)\frac{\partial\beta_i(T_1)}{\partial T_1}+2\Omega_i\tilde{\omega}_{ij}^\mathrm{xD}A^2_jA_i(T_1)\Big[\frac{1}{2}+\\
&+\frac{\cos(2\Omega_jT_0)}{2}\Big].
\end{split}
\]
Looking at the coefficients $F_1$ and $F_2$, we note that all the time dependent terms are associated with the slow time scale $T_1$ except for two terms oscillating with $T_0$. Since these two terms oscillate faster than all the other terms, they average to zero and we are left with
\[
\begin{split}
F_1&=2\Omega_i\frac{\partial A_i(T_1)}{\partial T_1}+\tilde{\Gamma}_iA(T_1)\Omega_i+\frac{\tilde{\gamma}_{ij}^\mathrm{xnl}}{2}A^2_jA_i(T_1)\\
F_2&=2\Omega_iA_i(T_1)\frac{\partial\beta_i(T_1)}{\partial T_1}+\Omega_i\tilde{\omega}_{ij}^\mathrm{xD}A^2_jA_i(T_1).
\end{split}
\]
To avoid secular terms, we need to impose $F_1=0$ and $F_2=0$. By satisfying these two conditions, first we find the time evolution of the displacement amplitude
\begin{equation}
    A_i(T_1)=ce^{-\left(\frac{\tilde{\Gamma}_i}{2}+\frac{\tilde{\gamma}_{ij}^\mathrm{xnl}}{4}A^2_j\right)T_1}
\end{equation}
where $c$ will be determined by the initial condition. The time evolution of $\beta_i$ is
\begin{equation}
    \beta_i(T_1)=\frac{1}{2}\tilde{\omega}_{ij}^\mathrm{xD}A^2_jT_1+\Phi.
\end{equation}
Since we are interested in the zero-th order solution, we obtain $u_i(t)$ substituting the expression for $\beta_i(T_1)$ and $A_i(T_1)$ in eq. (\ref{MS_eq11}) and going back to the initial time scale and initial parameters. The solution of the equation of motion is
\begin{equation}
    u_i(t)=\underbrace{A_{i,0}e^{-\left(\frac{\Gamma_i}{2}+\frac{\gamma_{ij}^\mathrm{xnl}}{4}A^2_j\right)t}}_{A_i(t)}\cos\underbrace{\left(\Omega_it+\frac{\omega_{ij}^\mathrm{xD}}{2}A^2_jt+\Phi\right)}_{\varphi_i(t)}\
\end{equation}
From the time evolution of the amplitude we observe an increment in the linear damping due to the cross nonlinear damping
\begin{equation}
    \Gamma_i'=\Gamma_i+\frac{\gamma_{ij}^\mathrm{xnl}}{2}A^2_j
\end{equation}
and from the time derivative of $\varphi(t)$ we can predict the frequency shift induced by the cross Duffing
\begin{equation}
    \Omega_i'=\Omega_i+\frac{\omega_{ij}^\mathrm{xD}}{2}A^2_j.
\end{equation}
Both the effects depend on the displacement amplitude square of the driven mode $A_j$.
\section{Perturbative derivation of Duffing parameter}
\label{app:Duffingsimulation}
%
Here we derive a general expression for the Duffing parameter, following simple energy considerations.
Assuming that the nonlinearities in our system are geometric by nature, the frequency shift of the Duffing resonator can be attributed to a non-negligible contribution of the elongation energy at large excitation amplitudes.
Given the total mechanical energy of the resonator, ${E_0=m_{\mathrm{eff}, i}\Omega_{i}^2A_{i,0}^2/2}$, the resonance frequency change can be approximated as follows 
\begin{align}
\Omega_i' = \Omega_{i}+\delta\Omega_{\mathrm{m}} &\approx \sqrt{\frac{2(E_0+E_\mathrm{nlin})}{m_{\mathrm{eff}, i}A_{i,0}^2}}\\
&\approx \Omega_{i}\left(1+\frac{E_\mathrm{nlin}}{2E_0}\right),
\end{align}
where $A_{i,0}$ is the amplitude of mode $i$.
Comparing this equation with that of the backbone curve
\begin{align}
\label{eq:DuffingElongEq1}
\Omega_i' = \Omega_{i}+\frac{3\alpha_i}{8m_{\mathrm{eff},i}\Omega_{i}}A^2_{i,0}
\end{align}
allows us to express the Duffing parameter in terms of the elongation energy
\begin{align}
\frac{E_\mathrm{nlin}}{m_{\mathrm{eff},i}\Omega_{i}A_{i,0}^2} &= \frac{3\alpha_i}{8m_{\mathrm{eff}, i}\Omega_{i}}A^2_{i,0}.
\end{align}
Following ref. \cite{Ugural2017}, the elongation energy can be expressed in terms of the displacement field profile ${w_i (x,y) = A_0\phi_i (x,y)}$ (with $i\in\left\{S,A\right\}$) as follows
\begin{align}
E_\mathrm{nlin} &= \frac{E}{8(1-\nu^2)}\int_V \left\{\left(\frac{\partial w}{\partial x}\right)^2+\left(\frac{\partial w}{\partial y}\right)^2\right\}^2 d\mathrm{V}.
\end{align}
In conjuction with eq. (\ref{eq:DuffingElongEq1}), we can thus express the Duffing parameter in terms of a volume integral of the normalized modeshapes $\phi_i$
\begin{align}
\label{eq:simpleDuffing}
\alpha_i = \frac{E}{3(1-\nu^2)}\int_V \left\{\left(\frac{\partial \phi_i}{\partial x}\right)^2+\left(\frac{\partial \phi_i}{\partial y}\right)^2\right\}^2 d\mathrm{V}.
\end{align}
This method is perturbative in that it uses the mode shapes obtained from solving the elastic equations that disregard the nonlinear energy terms.

As an example, we consider the Duffing parameter of a doubly-clamped beam, with sidelength $L$, thickness $h$ and width $w$.
The fundamental modeshape of the beam can be approximated as $\phi_\mathrm{beam}\approx\mathrm{sin}(\pi  x/L)$. Using eq. (\ref{eq:simpleDuffing}), we find that $\alpha=E h w\pi^4/(8(1-\nu^2)L^3)$.
Given that the effective mass of the beam is ${m_\mathrm{eff}=\rho h w L/2}$, the Duffing parameter can be expressed as ${\alpha=E m_\mathrm{eff}\pi^4/(4\rho(1-\nu^2)L^4)}$, in agreement with previous derivations of the Duffing parameter \cite{Unterreithmeier2009a}.
Following the same procedure, we obtain for the fundamental mode  $\phi_\mathrm{square}\approx \sin(\pi x /L)\sin(\pi  y/L)$ of a square membrane ${\alpha=5 E m_\mathrm{eff}\pi^4/(12\rho(1-\nu^2)L^4)}$.
\section{Finite element simulations}
\label{app:simulation}
%
As previously mentioned (cf. Appendix \ref{app:calibration}), our displacement calibration is reliant upon numerical simulations of the effective modal masses.
To this end, we use COMSOL Multiphysics to simulate the eigenmodes of the soft-clamped phononic dimer structures.
For all of our simulations we assume a film thickness of $14~\mathrm{nm}$, density $\rho=3200~\mathrm{kg/m^3}$, Young's modulus $E_1=270~\mathrm{GPa}$, Poisson's ratio of $\nu=0.27$ and initial stress ${\bar{\sigma}_{xx}=\bar{\sigma}_{yy}=1.27~\mathrm{GPa}}$.
Since the phononic crystal perforation results in a stress redistribution, we first simulate the new steady state distribution of in-plane tension, followed by an eigenfrequency analysis, yielding the eigenfrequencies and displacement profiles of the symmetric and antisymmetric modes.
Using the simulated displacement profiles, we perform volume integrals following eq. (\ref{eq:effMassDefinition}) for all modes of interest, yielding the effective masses for the respective modes (see Table \ref{tab:table-1} for a summary of modal masses).

In addition to the effective masses, we use the simulated eigensolutions to estimate the mechanical quality factors. Following \cite{Yu2012}, the total amount of dissipated elastic energy can be expressed in terms of the mode curvature
\begin{align}
\Delta W &= \int z^2 dz \iint \frac{\pi E_2(x,y)}{1-\nu^2} \set[\Bigg]{\bigg(\underbrace{\frac{\partial^2 w}{\partial x^2} + \frac{\partial^2 w}{\partial y^2}}_{\text{mean curvature}}\bigg)^2 \nonumber \\
\label{eq:fullDissipatedEnergyYu}
&\quad -2(1-\nu)\underbrace{\left(\frac{\partial^2 w}{\partial x^2}\frac{\partial^2 w}{\partial y^2}-\left(\frac{\partial^2 w}{\partial x\partial y}\right)^2\right)}_{\text{Gaussian curvature}}} dx dy,
\end{align}
where $E_2 (x,y)$ is the imaginary part of Young's modulus, as described by Zener's model (i.e., $E=E_1+i E_2$).
In our simulations we assume an isotropic loss tangent, and following \cite{Villanueva2014} we set $Q_\mathrm{int}=E_1/E_2=10^{2}~\mathrm{nm}^{-1}\times h = 1400$ (assuming a film thickness of $h=14~\mathrm{nm}$), amounting to $E_2 \approx 193~\mathrm{MPa}$.
Given the second order spatial derivatives in eq. (\ref{eq:fullDissipatedEnergyYu}), we ensure that the membrane geometries as densely meshed. 
Finally, the simulated displacement profiles are used to calculate the maximum kinetic energy, $W_{\mathrm{kinetic}} = \rho\Omega^2 \int w(x,y)^2 dV/2$, which, in conjunction with eq. (\ref{eq:fullDissipatedEnergyYu}), yields the mechanical quality factors for the symmetric and antisymmetric modes (Q=$2\pi W_{\mathrm{kinetic}}/\Delta W$).
We find that, on average, the Gaussian curvature results in approximately $10\%$ reduction of the quality factors. As such, the device performance is largely captured by the integrated mean modal curvature.
The simulated quality factors for all resonator geometries are listed in Table \ref{table:Qs}.

\begin{table}
\begin{tabular}{c|cc|cc|cc|cc}
$\theta$ & \multicolumn{2}{c}{0 unit cell} & \multicolumn{2}{c}{1 unit cell} & \multicolumn{2}{c}{2 unit cell} & \multicolumn{2}{c}{3 unit cell}\\
\hline
& $Q_\mathrm{S}$ & $Q_\mathrm{A}$ & $Q_\mathrm{S}$ & $Q_\mathrm{A}$ &$Q_\mathrm{S}$ & $Q_\mathrm{A}$ &$Q_\mathrm{S}$ & $Q_\mathrm{A}$\\
$0 \deg $& 147 & 137 & 142 & 137 & 140 & 140 & 138 & 140\\
$14 \deg$ & -- & -- & 142 & 143 & -- & -- & 140 & 139\\
$30 \deg$ & -- & -- & 136 & 147 & 137 & 142 & 138 & 141\\
$46 \deg$ & -- & -- & 142 & 143 & -- & -- & 140 & 139\\
\end{tabular}
\caption{\label{table:Qs} Simulated mechanical quality factors (units $10^6$) following eq. (\ref{eq:fullDissipatedEnergyYu}). Here we have assumed $E_2 \approx 193~\mathrm{MPa}$.
}
\end{table}

Finally, we use the simulated eigenmodes to estimate the Duffing parameters, following eq. (\ref{eq:simpleDuffing}).
We perform the spatial differentiations and volume integration within COMSOL, thus finding the self-Duffing parameters for the eigenmodes.
Our simulations predict self-Duffing parameters on the order of $\omega^\mathrm{sD}/2\pi\approx 2\times 10^{16}~\mathrm{Hz/m^2}$, in agreement with the measured values within the experimental scatter (cf. Section \ref{sec:nonlinearities}). For all but one dimer configuration (namely $0^\circ$ orientation and 0 unit cell separation) the symmetric and antisymmetric modes of a given structure have comparable self-Duffing parameters. For the device with the smallest dimer spacing, exhibiting the largest inter-defect coupling strength, we find that the simulated Duffing parameter of the antisymmetric mode is approximately twice as large as that of the symmetric mode. However, our experimental scatter is too large to test this prediction.

\section{Device fabrication}
\label{app:fabrication}
Our devices are fabricated on $500~\mathrm{\mu m}$ thick, 4" double-side polished silicon wafers. Following a low-pressure chemical vapour deposition of stoichiometric silicon nitride ($\mathrm{Si}_3\mathrm{N}_4$), we spin coat $1.5~\mathrm{\mu m}$ positive photoresist on the top (membrane) side of the wafer, followed by a softbake. A direct laser writer is used to write the phononic crystal pattern into the photoresist. Following development of the photoresist, we use an inductively coupled plasma (ICP) etch to transfer the pattern to the $\mathrm{Si}_3\mathrm{N}_4$ film. We repeat a similar lithographic and etch step for the backside of the wafer, in order to define rectangular openings in the silicon nitride. After stripping the photoresist, the wafers are placed inside a backside protecting PEEK holder, and etched in a potassium hydroxide (KOH) solution at $80^{\circ}$ for approximately 7 hours. The wafers are finally cleaned in a hot piranha solution and rinsed thoroughly in deionized water.

\begin{figure}[htb]
\centering
\includegraphics[width=1\linewidth]{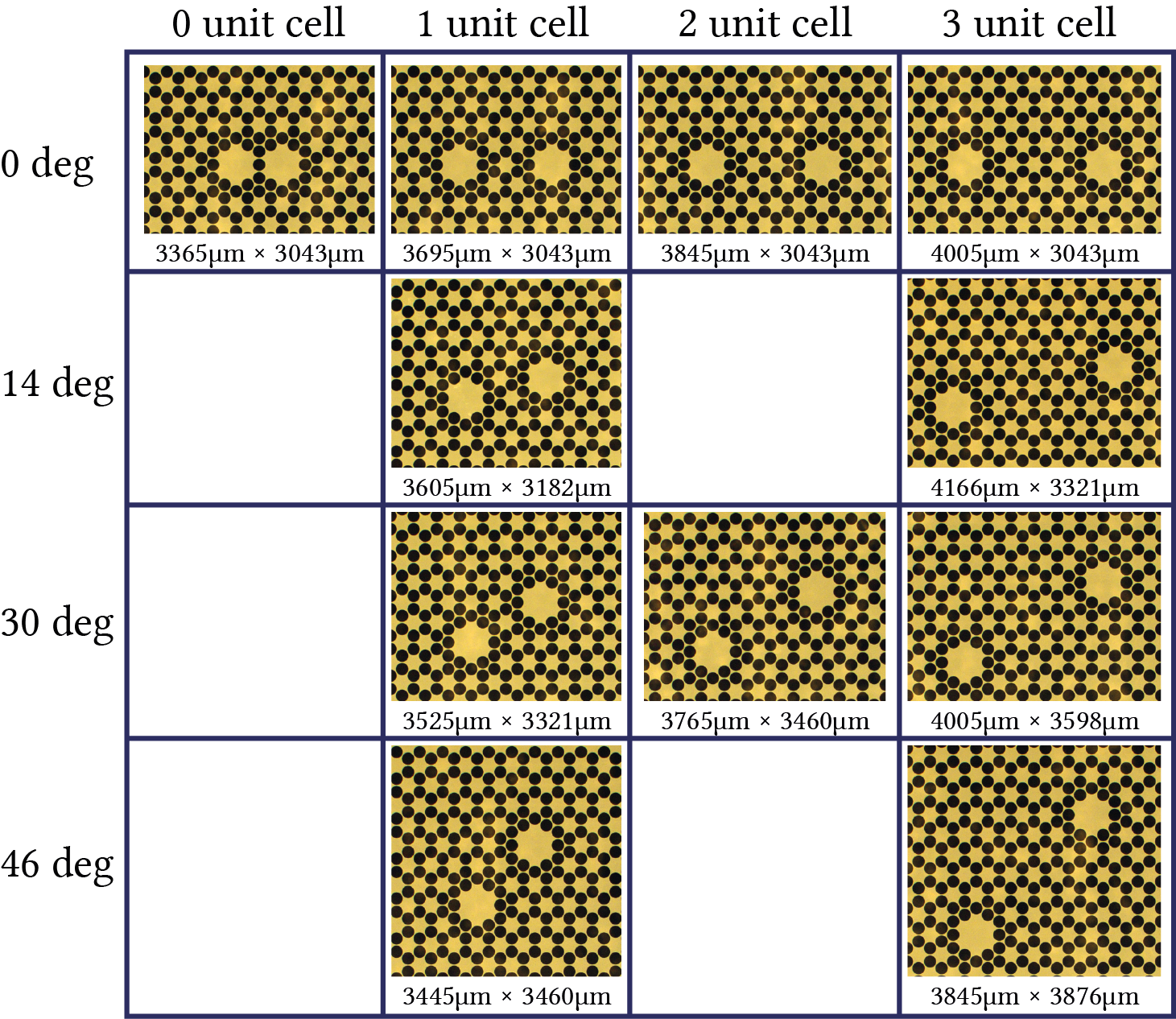}
\caption{Micrographs representative of all devices geometries characterized in this study. The outer dimensions of the individual devices are indicated below each micrograph. All devices have the same number of unit cells separating the individual defects and the silicon frame, resulting in a different overall membrane dimensions depending on the dimer orientation and defect separation.}
\label{f:devices}
\end{figure}

In Fig.~\ref{f:devices} an overview of all relevant device geometries is presented.


\end{document}